\PassOptionsToPackage{dvipsnames}{xcolor}
\documentclass[9pt,sigconf,nonacm,screen]{acmart}

\usepackage[all]{nowidow}
\usepackage{xcolor}
\usepackage{subcaption}
\usepackage{hyperref}
\usepackage{acronym}
\usepackage[capitalise,noabbrev]{cleveref}

\begin{document}

\author{Minghe Wang}
\email{mw@3s.tu-berlin.de}
\orcid{0009-0001-3780-5828}
\affiliation{%
    \institution{Technische Universit\"at Berlin}
    \department{Scalable Software Systems Research Group}
    \city{Berlin}
    \country{Germany}}

\author{Alexandra Kapp}
\email{ak@3s.tu-berlin.de}
\orcid{0000-0002-8348-8958}
\affiliation{
    \institution{Technische Universit\"at Berlin}
    \department{Scalable Software Systems Research Group}
    \city{Berlin}
    \country{Germany}}

\author{Trever Schirmer}
\email{ts@3s.tu-berlin.de}
\orcid{0000-0001-9277-3032}
\affiliation{
    \institution{Technische Universit\"at Berlin}
    \department{Scalable Software Systems Research Group}
    \city{Berlin}
    \country{Germany}}

\author{Tobias Pfandzelter}
\email{tp@3s.tu-berlin.de}
\orcid{0000-0002-7868-8613}
\affiliation{
    \institution{Technische Universit\"at Berlin}
    \department{Scalable Software Systems Research Group}
    \city{Berlin}
    \country{Germany}}

\author{David Bermbach}
\email{db@3s.tu-berlin.de}
\orcid{0000-0002-7524-3256}
\affiliation{%
    \institution{Technische Universit\"at Berlin}
    \department{Scalable Software Systems Research Group}
    \city{Berlin}
    \country{Germany}}

\renewcommand{\shortauthors}{Wang et al.}

\title{Exploring Influence Factors on LLM Suitability for No-Code Development of End User Applications}


\begin{abstract}
    No-Code Development Platforms (NCDPs) empower non-technical end users to build applications tailored to their specific demands without writing code.
    While NCDPs lower technical barriers, users still require \emph{some} technical knowledge, e.g., to structure process steps or define event-action rules.
    Large Language Models (LLMs) offer a promising solution to further reduce technical requirements by supporting natural language interaction and dynamic code generation.
    By integrating LLM, NCDPs can be more accessible to non-technical users, enabling application development truly without requiring \emph{any} technical expertise. 

    Despite growing interest in LLM-powered NCDPs, a systematic investigation into the factors influencing LLM suitability and performance remains absent. 
    Understanding these factors is critical to effectively leveraging LLMs capabilities and maximizing their impact. 
    In this paper, we investigate key factors influencing the effectiveness of LLMs in supporting end-user application development within NCDPs. 
    By conducting comprehensive experiments, we evaluate the impact of four key factors, i.e., model selection, prompt language, training data background, and an error-informed few-shot setup, on the quality of generated applications. 
    Specifically, we selected a range of LLMs based on their architecture, scale, design focus, and training data, and evaluated them across four real-world smart home automation scenarios implemented on a representative open-source LLM-powered NCDP.
    Our findings offer practical insights into how LLMs can be effectively integrated into NCDPs, informing both platform design and the selection of suitable LLMs for end-user application development.
\end{abstract}

\maketitle

\section{Introduction}
\label{sec:introduction}
Customizing applications enhances usability and effectiveness by delivering smart experiences tailored to users' unique needs and lifestyles.
For example, in smart home environments, some users may prefer window blinds that automatically adjust based on sunlight exposure to maintain steady indoor lighting, while others might prioritize wake-up routines synchronized with natural light or security systems that adapt to their daily schedules.
These personalized configurations not only enhance convenience and comfort but can also improve energy efficiency, security, and privacy~\cite{paper_werner2017_openhab,bao2016microservice}.

Different end users have unique preferences, routines, and priorities, which makes it difficult to capture in a one-size-fits-all application.
Achieving tailored functionality often requires coding knowledge, creating barriers for non-technical users.
No-code development platforms (NCDPs) have emerged as a solution, enabling non-technical users to create custom applications without requiring any technical expertise~\cite{blackstock2016fred}.
Although NCDPs significantly simplify the application development process, users still need to navigate complex interfaces and workflows, understand platform-specific components and abstractions~\cite{10.1145/3493369.3493600,10.1145/3366610.3368100,10.1145/3286719.3286725,10.1145/3286719.3286728}, which makes widespread adoption difficult, especially for those unfamiliar with structured automation concepts.
While rule-based platforms, e.g., IFTTT, reduce this barrier by providing basic automation, true interoperability across different ecosystems typically depends on API integration and programming skills.
This reliance on technical skills further limits accessibility to non-technical users.

For this limitation, Large Language Models (LLMs) present new opportunities for lowering the technical knowledge requirement for application development~\cite{llm4faas}.
By leveraging natural language processing and code generation capabilities, LLMs can transform user intentions into functional codes~\cite{llm4faas,monteiro2025nocodegpt} - a natural language description of desired functionality is the only information required from the end users.
However, variations in architecture, capabilities, training data, and design objectives among different LLMs can result in performance differences across tasks thus influence their suitability for applications development.
While \emph{leaderboards} compare models on general purpose tests regarding~\cite{artificialanalysis,llmstats,huggingfaceopenllm,chatbotarena}, e.g., knowledge, reasoning, or code generation, a more detailed comparison is necessary to understand the influence of varying model factors for specific tasks. 
For instance, no-code development of applications by non-technical end users will have completely different demands on LLMs than code generation as part of a professional developer's toolchain. 
To optimize NCDP development and broaden access, it is critical to understand LLM performance differences in translating natural language instructions into code for user-driven application customization tasks.

While the landscape of LLM-powered NCDPs is rapidly evolving, existing research has primarily focused on the feasibility of LLM integration~\cite{cai2023low, chen2024llm2automl, hagel2024turning, gao2024chatiot,llm4faas}.
Systematic, comparative studies of LLM performance within these platforms remain scarce.
Rather than introducing platform-level variability, this paper focuses on isolating and systematically analyzing the factors that influence the suitability of LLMs for end-user application development on NCDPs.
Since LLMs serve as the core and shared component across these LLM-powered NCDPs, studying their behavior allows us to derive broadly applicable, generalizable, and transferable insights.
Our study investigates three critical aspects, i.e., LLM selection, prompt language, and original model training data.
We conduct our evaluation using a representative LLM-powered NCDP~\cite{llm4faas}, designed specifically for end users without technical backgrounds, and build on Function-as-a-Service (FaaS) abstractions.
This architecture decouples infrastructure handling concerns from function logic generation, which has been shown to enhance the reliability of LLM outputs, making it server as a stable and representative foundation for isolating and examining LLM-specific factors in NCDPs.
We evaluate eight LLMs on four smart home automation tasks based on real non-technical user description and investigate the following:
\begin{enumerate}
    \item Impact of LLM choice on performance across varying task complexities, examining whether certain models are better suited for development tasks with non-technical user interactions.
    \item The robustness of expectable NCDP performance for different languages, namely, Chinese and English.
    \item The role of LLM linguistic backgrounds which the text corpora a model trained on, in shaping usability and effectiveness of NCDPs.
\end{enumerate}

By evaluating these factors and incorporating a syntactic error-informed few-shot approach, we aim to provide insights into the design and optimization of future LLM-powered NCDPs, making application development more accessible and enabling non-technical users to fully harness the potential of end-user applications.

The rest of the paper is organized as follows:
\Cref{sec:background} gives an overview of the concepts of LLMs, NCDPs, and FaaS paradigm; it also presents related work.
\Cref{sec:methodology} describes the methodology and study design we followed while investigating the impact factors of LLM-powered NCDPs, and
\cref{sec:results} presents the experiment results and findings.
Finally, we critically discuss our findings in \cref{sec:discussion} before coming to a conclusion in \cref{sec:conclusion}.

\section{Background and Related Work}
\label{sec:background}
LLMs and NCDPs are revolutionizing the development and deployment of applications by simplifying the design process and enhancing user accessibility, while FaaS further streamlines automation by enabling scalable, event-driven task execution.

\subsection{Large Language Models}
\label{sec:background:llm}

LLMs are advanced artificial intelligence models trained on vast amounts of text data to understand, process and generate human-like responses.
A wide range of LLMs, varying in architecture, scale, training data, and deployment context, become available, from proprietary models served via APIs to open-source models optimized for local or custom deployment, e.g., GPT series from OpenAI~\cite{achiam2023gpt}, Gemini from Google~\cite{team2023gemini}, and LLaMA from Meta~\cite{dubey2024llama}, etc.
These models use deep learning techniques, particularly transformer architectures, to analyze and predict text patterns, making them highly effective for various natural language processing tasks, e.g., translation, summarization, and providing contextually relevant responses.
Their ability of translating natural language into structured logic and automating tasks makes LLMs particularly beneficial for NCDPs, enabling non-technical users to interact with systems through conversational prompts~\cite{chen2024llm2automl,hagel2024turning,ulfsnes2024transforming, kok2024iot}.

Essentially, integrating LLMs into the NCDPs, the barrier that end users face in customized software development is significantly lowered, as users can generate functional code, debug issues, and optimize application logic via intuitive, natural language based interfaces~\cite{cai2023low,paliwal2024low, martins2023combining,gao2024chatiot}.

\subsection{No-Code Development Platforms}
\label{sec:background:no-code}

NCDPs enable users to design, deploy, and modify applications through visual interfaces and pre-built drag-and-drop components, eliminating the need for coding expertise~\cite{10.1145/3493369.3493600,10.1145/3366610.3368100,10.1145/3286719.3286725,10.1145/3286719.3286728,el2023low}.
The NCDPs democratize software development by allowing individuals without programming knowledge to build functional applications efficiently.
This approach streamlines the development process, leading to fast deployment, reducing costs, and enabling flexibility~\cite{upadhyaya2023low,sufi2023algorithms,chen2022devicetalk}.

The emergence of LLMs has opened new avenues for enhancing NCDPs. 
With LLM-powered NCDPs, non-technical users can describe their desired functionalities in natural language, LLM either generate the intermediate artifacts or provide functional codes to support the application development~\cite{martins2023combining, gao2024chatiot,llm4faas,monteiro2025nocodegpt}.

\subsection{Function-as-a-Service(FaaS) Paradigm}
\label{sec:background:faas}

Function-as-a-Service (FaaS) is a serverless computing model that allows developers to deploy individual functions that respond to specific events or triggers, without the need to manage servers or underlying infrastructure~\cite{paper_bermbach2021_cloud_engineering,paper_pfandzelter2019_functions_vs_streams}.
As FaaS also comes with a simple programming model, i.e., functions are typically stateless, small, and event-driven with the provider handling all operational complexity, integrating it with LLMs can  significantly reduce the task complexity that LLMs have to handle when generating custom applications based on natural language prompts of non-technical users~\cite{llm4faas}.
As a result, this combination has the promise to enhance and expand the current landscape of NCDP.

\subsection{Related Work}
\label{sec:background:related-work}
Prior work has examined the feasibility of integrating LLMs into NCDPs (\cref{sec:background:related-work:feasibility}) and the factors affecting the suitability of LLMs from technical perspectives (\cref{sec:background:related-work:suitability}).

\subsubsection{LLM Feasibility in NCDPs}
\label{sec:background:related-work:feasibility}
Existing work investigating the feasibility of LLMs in NCDPs varies in terms of targeted application domains, employed technical approaches, and underlying design paradigms.
Some work focuses on translating natural language into structured workflows or applications, for example, 
Cai et al.~\cite{cai2023low} present an LLM-driven low-code approach that generates editable flowcharts from user instructions for iterative refinement; 
Esashi et al.~\cite{esashi2025action} target FaaS workflow generation to benefit Cloud developers, using a dataset for compositional multi-tool tasks, though excluding execution; 
and Wang et al.~\cite{llm4faas}, our prior work, use \emph{GPT-4o} to generate serverless functions and deploy them via FaaS platform, thereby enabling customized application development for non-technical users.
There is also work aiming to improve LLMs' generation quality through multiagent architectures or domain-specific prompting, for example,
Gao et al.~\cite{gao2024chatiot} decompose IoT trigger-action-program creation into subtasks coordinated by specialized agents, and Koubaa et al.~\cite{koubaa2025next} integrate ontologies into prompts for Robot Operating System (ROS2) specific command generation.
Some work adapts LLM-based platforms to particular domains or examining broader impacts, for example,
Monteiro et al.~\cite{monteiro2025nocodegpt} propose NoCodeGPT for web application development with rollback mechanisms,
Hagel et al.~\cite{hagel2024turning} automate the DSL models generation based on technical descriptions,
Chen et al.~\cite{chen2024llm2automl} introduce a template-bounded no-code automated machine learning (AutoML) framework,
and Liu et al.~\cite{liu2024empirical} conduct an empirical comparison of traditional low-code programming and an LLM-based approach using GPT3, by analyzing developer discussions on Stack Overflow.

These work confirm the feasibility of embedding LLMs in NCDPs for diverse scenarios, however, they typically assess only a single model, focusing on whether an approach works.
Beyond feasibility, the underlying factors that can affect system performance are not investigated in these studies yet, leaving it unclear whether the presented feasibility would generalize to other models or conditions.
This remains a critical next step question in LLMs' suitability for no-code or low-code development, i.e., what LLM-related factors affect the performance of LLM-powered no-code and low-code platforms.

\subsubsection{LLM Suitability in other Domains}
\label{sec:background:related-work:suitability}

There exists research examining LLM suitability in diverse technical domains, for example, Li et al.~\cite{li2025large} compare GPT models with smaller fine-tuned models for detecting self-admitted technical debt, proposing a hybrid framework to balance precision and recall; 
Lu et al.~\cite{lu2025performance} investigate distributed training strategies for large Transformer models, analyzing performance-memory trade-offs across model architectures and parallelization methods; 
Petrukha et al.~\cite{petrukha2025swifteval} introduce SwiftEval, a benchmark for assessing LLMs on native Swift programming tasks, revealing performance variations across model families and sizes.

These studies show that understanding why an LLM performs well is crucial for informed adoption, however, they focus on technical workflows and users, whose interaction patterns and error tolerance differ from NCDP contexts.
Designed for non-technical users, NCDPs aim to minimize required technical operations, introducing distinct usability needs and interaction constraints, underscoring the importance of examining LLM suitability specifically in NCDPs, where effectiveness may depend on factors beyond those studied in technical domains.

\section{Methodology}
\label{sec:methodology}
To investigate the factors shaping LLM-powered NCDPs, we introduce our methodology from four aspects: base platform selection (\cref{sec:methodology:llm4faas}), experimental design (\cref{sec:methodology:experimental_design}), dataset construction (\cref{sec:methodology:dataset}), and evaluation metrics (\cref{sec:methodology:metrics}).
Specifically, our experiments focus on four key aspects, i.e., model selection, variation in prompt language, community background of LLMs, and the impact of a runtime syntactic error informed few-shot setting.

\subsection{Base Platform Selection}
\label{sec:methodology:llm4faas}
To assess the impact of LLMs on no-code development, we select a base platform, \emph{LLM4FaaS}, to perform evaluation.
Specifically, \emph{LLM4FaaS} leverages the high levels of abstraction of FaaS paradigm to handle code execution and operation, enabling LLMs a sole focus on core functionality generation.
As we focus on the LLM-related factors, we expect the base platform with a clear architecture so that the LLM behavior can be isolated and easily observed.
Notably, we do not perform a comparative analysis of different NCDPs, and discuss this in \cref{sec:threats:platform}.
We reuse and extend the \emph{LLM4FaaS} dataset to conduct the experiments and also discuss the dataset choice in \cref{sec:threats:dataset}.

\subsection{Experiment Plan}
\label{sec:methodology:experimental_design}
To systematically evaluate the impact of LLM selection for NCDPs, we conduct experiments considering multiple aspects.  
We select five LLMs with different strong suits in, i.e., design focus, model size, domain-specific optimization, to evaluate and compare their performance in terms of syntactic and semantic success of code production.
Additionally, as the user inputs can influence the understanding of LLMs, we evaluate the model performance with two different user input languages, i.e., in Chinese and English.
Also, as LLMs trained in different linguistic environments may exhibit varying performance across languages, we explore the LLMs performance by evaluating three mainstream Chinese LLMs.
Finally, we explore the impact of zero-shot and few-shot settings to LLM-based NCDPs.

\subsubsection{Model Selection}
\label{sec:methodology:experiment:model}
With the rapid proliferation of LLMs, it is essential to evaluate their feasibility within NCDPs.
A thorough analysis of various LLMs can help optimize model selection strategies, ensuring these platforms achieve peak performance, efficiency, and user-friendliness.
We intentionally select models that vary in architecture, scale, and domain specialization to investigate how fundamental design properties affect model behavior. 
In this way, we can derive insights that are more robust and generalizable, beyond the short-term performance of any single model version.

To explore this, we compare five mainstream LLMs from different aspects of consideration, i.e.,
\begin{enumerate}
    \item \emph{GPT-4o}, a general purpose model with advanced capacity.
    \item \emph{GPT-4o-mini}, a resource-efficient and cost-effective model, to assess if a lightweight alternative can match the performance of larger models.
    \item \emph{Copilot}, a model optimized for software development, to examine whether a domain-specialized LLM outperforms general-purpose ones.
    \item \emph{LLaMA}, an open-source model, to evaluate the feasibility of self-hosted or fine-tuned LLMs for customization and flexibility.
    \item \emph{Gemini}, a model with strong reasoning and multimodal capabilities, to assess its effectiveness in handling complex prompts and broader contextual understanding.
\end{enumerate}
This comparison provides insights into the trade-offs between model size, design focus, and domain-specific optimization, guiding future adoption strategies for LLM-powered NCDPs.

\subsubsection{Prompt Language}
\label{sec:methodology:experiment:prompt}

The input language of LLMs can presumably have a large impact depending on the text corpora a model is trained on.
Models developed in English-speaking countries, i.e., English models, will presumably be trained with more English texts while models developed in from Chinese developer teams, i.e., Chinese models, may handle Chinese user input better.
To assess this language-specific impact, we conduct an experiment examining how prompt language, i.e., English and Chinese, affects LLMs performance.

\subsubsection{Community Background}
\label{sec:methodology:experiment:community}
The consideration of prompt language choice also underscores the importance of evaluating the performance with Chinese LLMs.
We select three mainstream Chinese LLMs to evaluate their performance, i.e., \emph{Alibaba Qwen}, \emph{Baidu Qianfan} and \emph{DeepSeek R1}, which also differ in design focus.
\begin{enumerate}
    \item \emph{Alibaba Qwen}, a model leveraging a Mixture-of-Experts (MoE) architecture, which is good at handling long-context tasks and large-scale language understanding.
    \item \emph{Baidu Qianfan}, a search-integrated LLM model optimized for Chinese natural language processing and enterprise applications.
    \item \emph{DeepSeek R1}, a reasoning-focused model, which incorporates chain-of-thought (CoT) prompting, excelling in domains, e.g., mathematics, programming, and complex multistep problem-solving.
\end{enumerate}

By comparing the performance of Chinese and English LLMs, we investigate whether aligning an LLMs training corpus with the input language improves its performance of LLM-powered NCDPs.

\subsubsection{Few-shot vs. Zero-shot Experiment Setting}
\label{sec:methodology:zero_shot_few_shot}
To assess the impact of prompting strategies on the performance of LLM-based NCDPs, we additionally conduct a few-shot experiment and compare it with the default zero-shot setting. 
Specifically, we design a feedback-based few-shot experiment leveraging runtime syntactic error as dynamic, task-specific guidance for iterative code refinement.
This approach avoids the potential noise and bias introduced by irrelevant examples, which are difficult to predefine given the highly customized and user-specific nature of NCDP tasks, especially for non-technical users.

\begin{itemize}
    \item \textbf{Zero-shot setting:} The LLMs generate code based solely on a structured prompt that contains the user requirements, without access to prior outputs or error feedback.
    \item \textbf{Few-shot setting:} The LLM receives additional context in the form of runtime error messages from previous attempts. 
    These serve as feedback to guide subsequent code refinement. 
    The generation process is allowed to iterate up to three times.
\end{itemize}

The few-shot setup aims to simulate a realistic development scenario by iteratively refining code with the feedback of interpreter or compiler, providing insights into the self-correct ability of LLM. 
We restrict the experiment to syntactic errors, which yield objective and consistent outcomes.
In contrast, semantic error handling typically requires clarification of user intent, which is inherently subjective, particularly in NCDP contexts involving non-technical users, and thus infeasible for standardized evaluation.

\subsection{Dataset}
\label{sec:methodology:dataset}
We use and extend an existing dataset of 26 real users natural language description for 4 smart home automation tasks in varying levels of complexity\cite{llm4faasdataset}.
Specifically, the complexity of task from simple to complex varies as follows, and we present an example in \cref{fig:methodology:user-response-example}.
\begin{enumerate}
    \item \emph{Simple}: a use case which expects a straightforward device control functionality.
    \item \emph{Medium}: introduce an additional layer of complexity, where the system is expected to handle three keyword-based or time-based automation sub-tasks.
    \item \emph{Advanced}: the scenario complexity improves by involving three sub-tasks and triggering smart devices based on the real-time sensor readings. 
    \item \emph{Complex}: the most complex scenario, where the complexity arises from the necessity of considering device coordination, balancing user behavior with environmental conditions, and accounting for the potential diversity and uncertainty of user preferences, all simultaneously
\end{enumerate}

To enhance dataset coverage and support our experiments, we extend it in the following ways:

\begin{figure*}
    \centering
    \includegraphics[width=\linewidth]{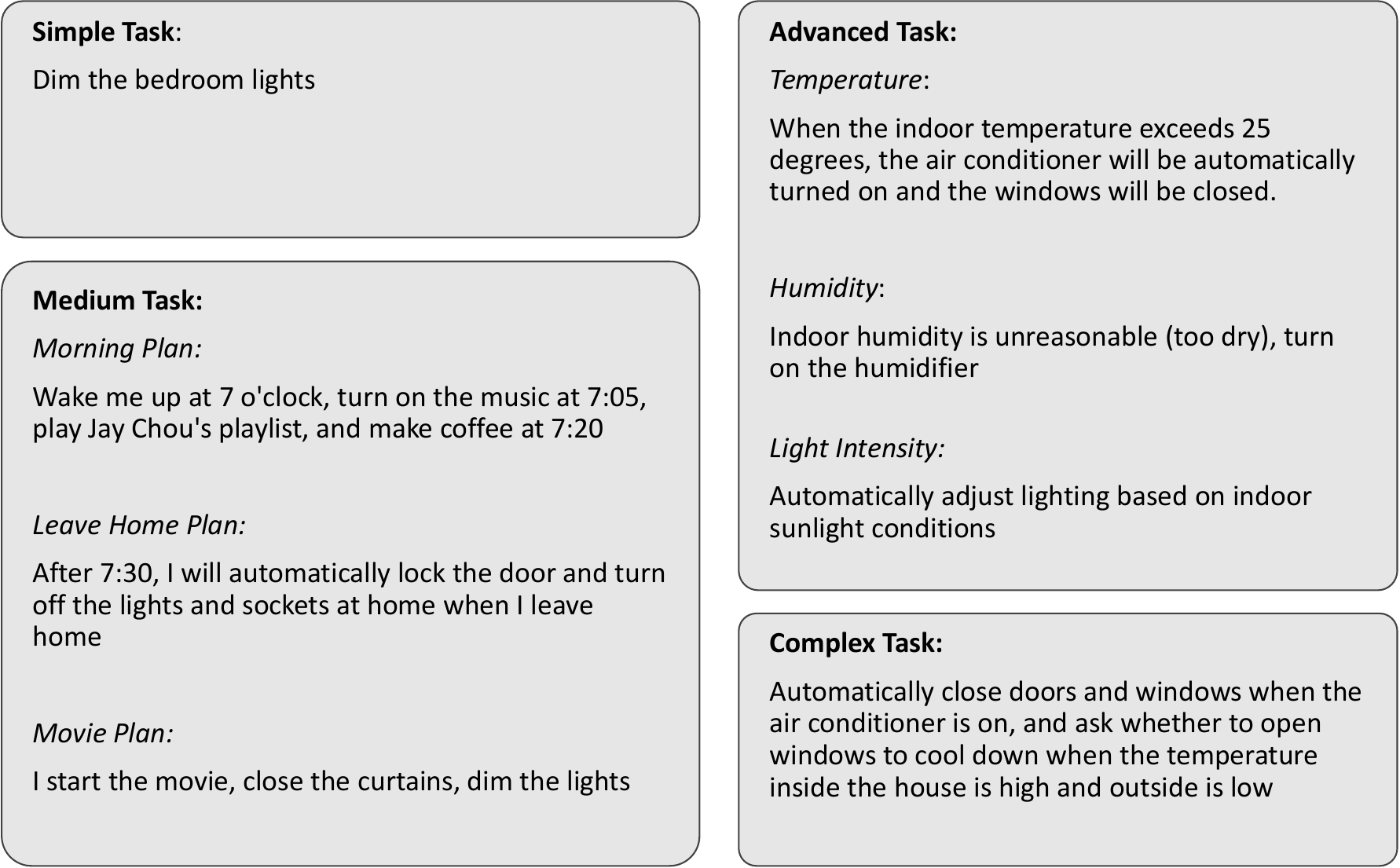}
    \caption{User Response to Tasks of Varying Complexity: 
        This figure presents a real user response from the dataset, illustrating four smart home automation tasks across different complexity levels. 
        The response showcases increasing complexity, from simple device control to intricate automation involving real-time sensor data and user-environment coordination. 
        The original response is in Chinese, and we provide an English translation for clarity.}
    \label{fig:methodology:user-response-example}
\end{figure*}

\begin{figure*}
    \centering
    \includegraphics[width=\textwidth]{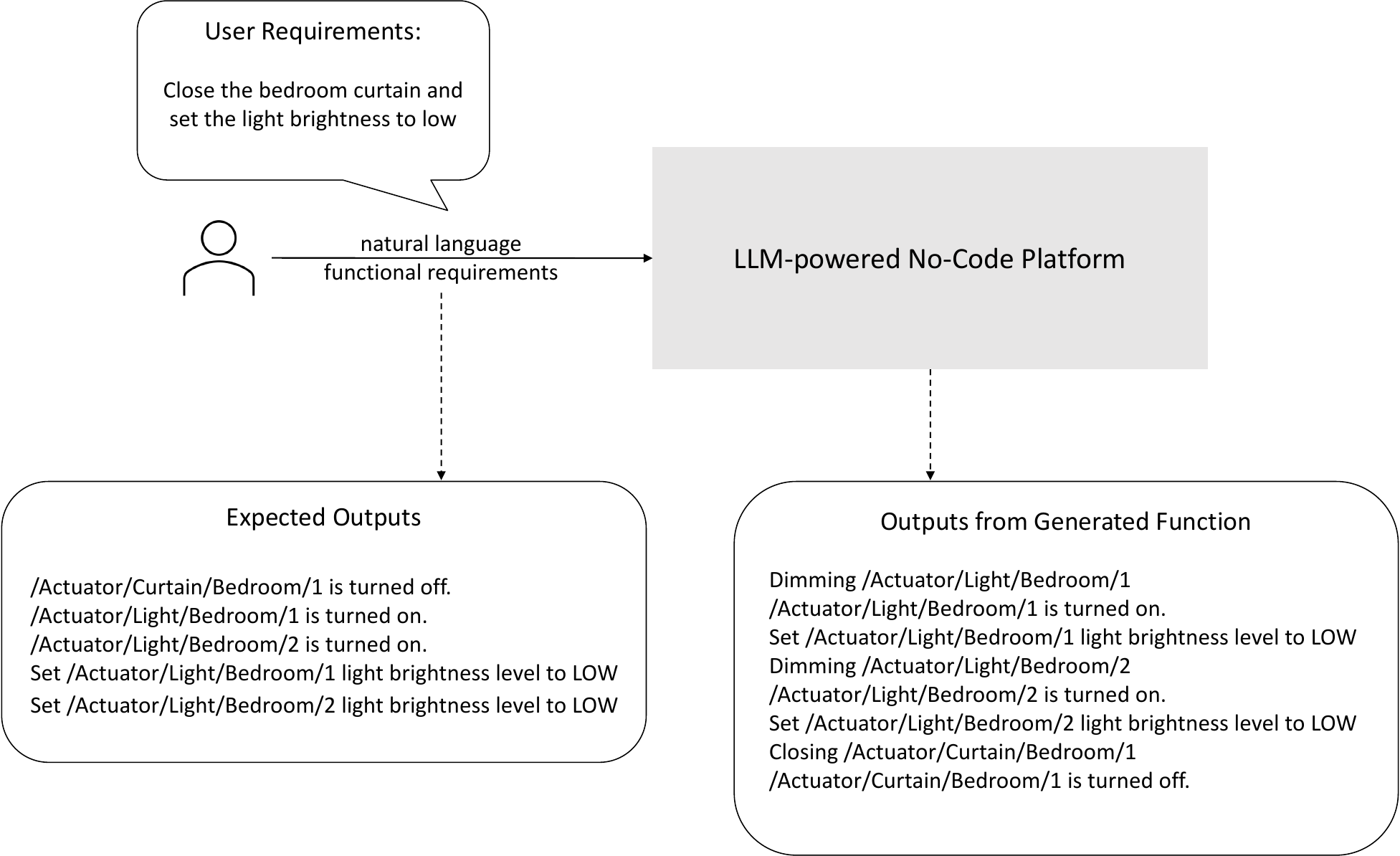}
    \caption{Dataset Example: We use a LLM-powered No-Code Platform, i.e., \emph{LLM4FaaS}, as the base NCDP for evaluation.
        Specifically, it takes the user natural language description in combined with the project context and reference code as input to LLMs.
        To evaluate the accuracy of LLM-generated results, we set the ground truth based on the user requirements.
        The accuracy is set by comparing the output from generated results and the ground truth.
        Essentially, the dataset consists of user requirements and the corresponding ground truth.}
    \label{fig:methodology:evaluation-workflow}
\end{figure*}

\subsubsection{Input Language Variation}  
To assess the impact of input language on LLMs performance, we translate the original Chinese user responses into English using Google Translate~\cite{googletranslate}, as English is the primary language for most LLMs. 
This allows for a direct comparison of LLM outputs across different input languages, i.e., Chinese and English, revealing potential language biases and inconsistencies.

\subsubsection{Ground Truth Definition}  
To evaluate the performance of LLM-generated results, we define the ground truth, constructed based on the expected console output, which is derived from user responses and the intended system response.
The ground truth is determined on a per-user and per-task basis, representing the expected system response for each automation task from every user.  
This ensures consistency in evaluation and allows for an objective comparison between LLM-generated outputs and the expected results. 

Essentially, our dataset consists of user natural language descriptions of smart home automation tasks, the translated English version, and ground truth outputs serving as evaluation baseline.
We show an example of dataset usage in \cref{fig:methodology:evaluation-workflow}.

\subsection{Evaluation Metrics}
\label{sec:methodology:metrics}
We first evaluate the \emph{syntactic success} and \emph{semantic success} rate of LLM-generated results to assess the quality of the generated code.
Specifically, the \emph{syntactic success} is defined as the absence of errors in the generated results, and the \emph{semantic success} is defined as a complete (100\%) alignment with user requirements.
While these metrics can showcase the feasibility, this binary classification fails to capture varying degrees of semantic accuracy, potentially overlooking cases where the results are partially aligned with user intent.

To address this limitation, we refine the evaluation metrics that quantify \emph{semantic accuracy} on a continuous scale, allowing for a more granular assessment of LLM-generated outputs.
In detail, the semantic accuracy rate use coverage match rate (CMR) to measure the extent to which the ground truth (GT) outputs are successfully covered by the LLM-generated outputs (MO), allowing for variations in wording and additional non-disruptive information in MO.
CMR quantifies the proportion of GT entries that are correctly identified within MO. It is formally defined as:

\begin{equation}
    CMR = \frac{|C|}{|GT|}
\end{equation}

, where:
\begin{itemize}
    \item \( |C| \) represents the number of GT entries successfully found in MO.
    \item \( |GT| \) denotes the total number of entries in the ground truth set.
\end{itemize}

A CMR of 100\% indicates that all GT outputs are fully covered in MO, i.e., achieves \emph{semantic success}, while lower scores suggest missing or incorrectly generated outputs.
For example, if the light should be turned on and the curtain closed, but only one task is accomplished, this results in 50\% \emph{semantic accuracy}.
This metric provides a more flexible assessment of model performance in real-world applications where exact textual matches may not always be required, but semantic correctness is essential.

Here, we allow the existence of additional outputs, e.g., description information before triggering the smart devices, as long as the core functionality is correctly implemented.
While we acknowledge that additional outputs in LLM-generated results may introduce noise, our current evaluation focuses solely on coverage accuracy.
Thus, we do not incorporate a separate noise rate metric in this study.

Our three metrics, i.e., \emph{syntactic success}, \emph{semantic success}, and \emph{semantic accuracy}, are conceptually aligned with standard evaluation metrics such as precision, recall, and F1-score, but adapted for NCDPs.
Specifically, \emph{syntactic success} checks whether generated code runs without errors, i.e., a prerequisite for meaningful evaluation.
\emph{Semantic success} is analogous to \emph{precision}: the proportion of syntactically valid outputs that fully satisfy the task requirements.
\emph{Semantic accuracy} is analogous to \emph{recall}: the degree to which the task requirements are satisfied, even partially, by valid outputs.
Taken together, \emph{semantic success} and \emph{semantic accuracy} jointly reflect both correctness and completeness, thus serving a role similar to the \emph{F1-score} in capturing overall task-level performance.
We adopt these tailored metrics to better reflect utility and reliability in no-code settings.

\section{Results}
\label{sec:results}
We present the experimental results, analyzing the impact of key factors on the performance of LLM-powered NCDPs.
The results highlight key findings across model selection (\cref{sec:results:model_selection}), prompt language (\cref{sec:results:prompt_language}), community background (\cref{sec:results:community}), and zero-shot vs. few-shot settings (\cref{sec:results:zero_shot_few_shot}), providing insights into their respective impacts (\cref{sec:results:findings}).

\subsection{Model Selection}
\label{sec:results:model_selection}

\begin{figure*}
    \centering
    \begin{subfigure}{0.49\linewidth}
        \centering
        \includegraphics[width=\textwidth]{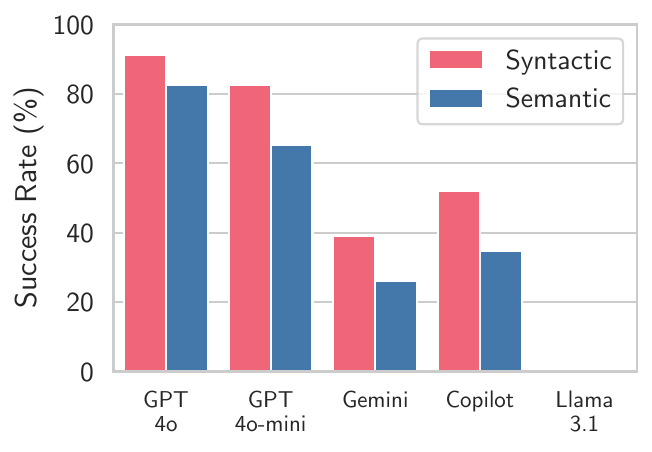}
        \caption{Easy}
        \label{fig:mainstream-easy}
    \end{subfigure}%
    \hfill
    \begin{subfigure}{0.49\linewidth}
        \centering
        \includegraphics[width=\textwidth]{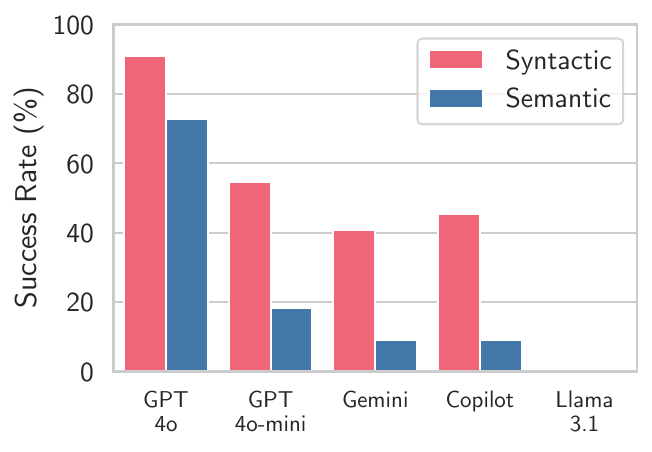}
        \caption{Medium}
        \label{fig:mainstream-medium}
    \end{subfigure}%
    \vfill
    \begin{subfigure}{0.49\linewidth}
        \centering
        \includegraphics[width=\textwidth]{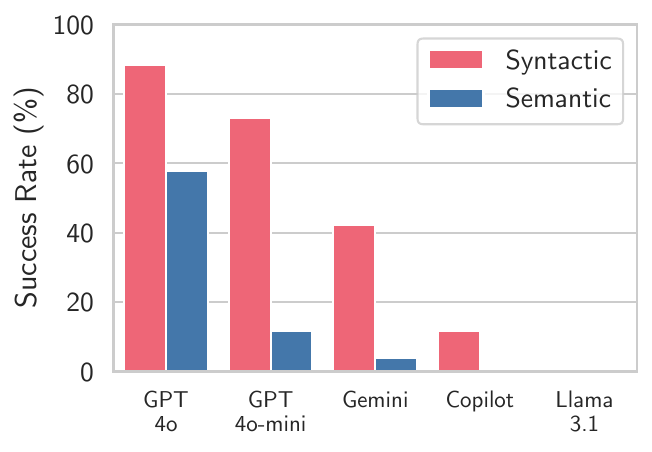}
        \caption{Advanced}
        \label{fig:mainstream-advanced}
    \end{subfigure}%
    \hfill
    \begin{subfigure}{0.49\linewidth}
        \centering
        \includegraphics[width=\textwidth]{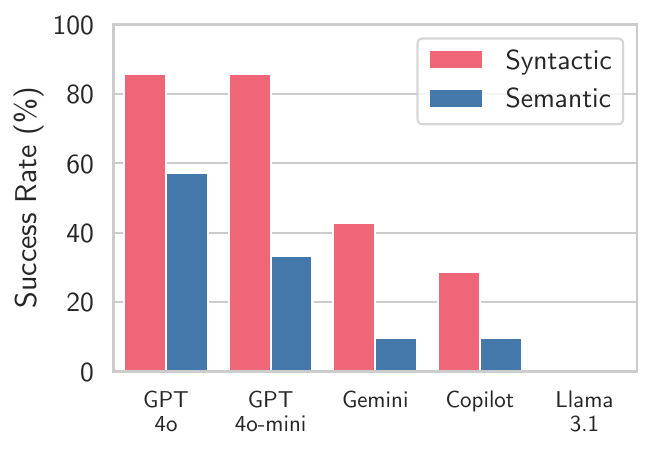}
        \caption{Complex}
        \label{fig:mainstream-complex}
    \end{subfigure}%
    \caption{We consider syntactic success as error-free results and semantic success as results fully meets user requirements.
        The graphs depict results based on Chinese user prompts, the original language of the used dataset.
        \emph{GPT-4o} shows distinct advantages in both syntactic and semantic success compared to other models.
        \emph{GPT-4o-mini} performs adequately on the easy task, but the performance drops significantly on more complex ones.
    }
    \label{fig:mainstream-semantic-syntactic}
\end{figure*}

\begin{figure*}
    \centering
    \begin{subfigure}{0.49\linewidth}
        \includegraphics[width=\textwidth]{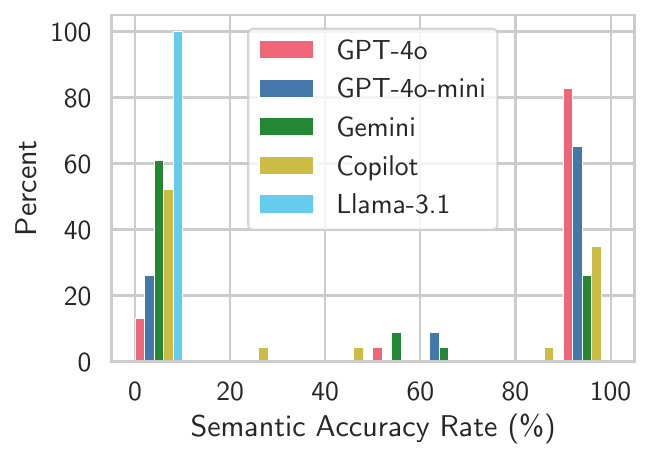}
        \caption{Easy}
        \label{fig:result:zh-count-easy}
    \end{subfigure}
    \hfill
    \begin{subfigure}{0.49\linewidth}
        \includegraphics[width=\textwidth]{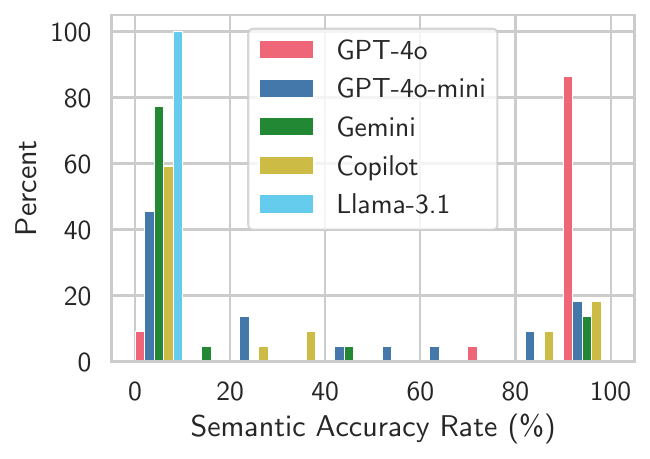}
        \caption{Medium}
        \label{fig:result:zh-count-medium}
    \end{subfigure}
    \hfill
    \begin{subfigure}{0.49\linewidth}
        \includegraphics[width=\textwidth]{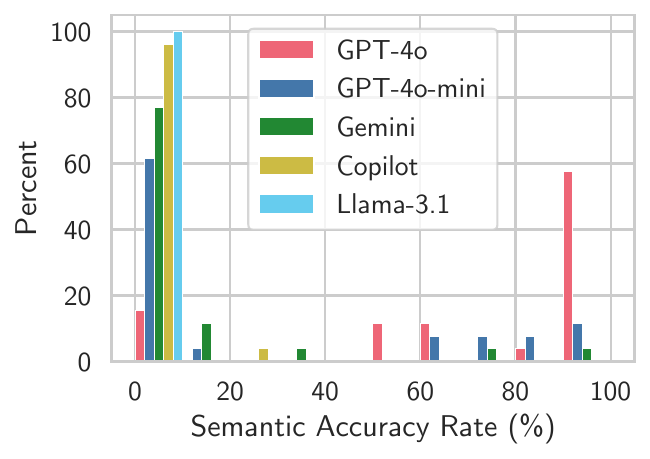}
        \caption{Advanced}
        \label{fig:result:zh-count-advanced}
    \end{subfigure}
    \hfill
    \begin{subfigure}{0.49\linewidth}
        \includegraphics[width=\textwidth]{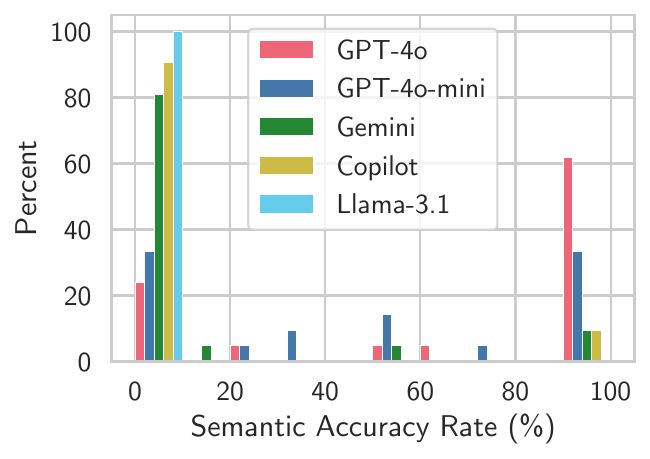}
        \caption{Complex}
        \label{fig:result:zh-count-complex}
    \end{subfigure}

    \caption{
        \emph{GPT-4o} consistently outperforms other models in semantic accuracy across tasks of varying difficulty, showing a higher accuracy rate in cases of non-semantic-success. 
        As the task complexity increases, the semantic accuracy distribution of results becomes more dispersed.
        The results show a bimodal distribution, with most tasks either failing completely or achieving 100\% success.
    }
    \label{fig:result:mainstream-semantic-count}
\end{figure*}

We compare the performance variance across five mainstream models, i.e., \emph{GPT-4o}, \emph{GPT-4o-mini}, \emph{Gemini-1.5-flash}, \emph{LLaMA-3.1}, and \emph{Copilot}.
First, we show the \emph{syntactic and semantic success rate} of results in \cref{fig:mainstream-semantic-syntactic}, then illustrate the \emph{semantic accuracy} distribution in \cref{fig:result:mainstream-semantic-count}.

The ranking of model performances for all or most experiments remains stable in the following order: \emph{GPT-4o}, \emph{GPT-4o-mini}, \emph{Gemini-1.5-flash}, \emph{Copilot}, and \emph{LLaMA-3.1}.
Among these, \emph{GPT-4o} and \emph{GPT-4o-mini} show the best performance in both average syntactic and semantic success rate among all tasks.
\emph{Gemini} and \emph{Copilot} have a similar performance for average semantic and syntactic success rate, where \emph{Copilot} performs better on simple tasks, and \emph{Gemini} in complex tasks.
\emph{LLaMA} fails to generate functions with all the user prompts.
Specifically, the average syntactic success rate among all task complexity levels of \emph{GPT-4o} and \emph{GPT-4o-mini} is 89.10\% and 73.99\%, respectively, while the other LLMs are below 50\%. 
The average semantic success rate of \emph{GPT-4o} among all tasks is 67.54\%, whereas the second best, \emph{GPT-4o-mini}, has dropped to 32.07\%, and other LLMs is below 15.00\%.

In addition to the semantic success rate, we are also curious about how close the results are to the semantic success, given by the \emph{semantic accuracy rate}.
Figure \ref{fig:result:mainstream-semantic-count} depicts the semantic accuracy, showing a high-level of semantic accuracy of \emph{GPT-4o}, which still demonstrates a clear advantage over the competing models across tasks in all difficulty levels.
For all the models, the distribution of semantic accuracy rate becomes more dispersed as the task complexity increases.
Notably, the results of complex tasks sometimes exhibit higher semantic success than those of medium and advanced tasks, i.e., it does not always align with the predefined task complexity.
This may be due to the fact that both medium and advanced tasks involve three unrelated sub-tasks, which can lead to more semantic failures.

\subsection{Prompt Language}
\label{sec:results:prompt_language}

\begin{figure*}
    \centering
    \begin{subfigure}{0.49\linewidth}
        \centering
        \includegraphics[width=\textwidth]{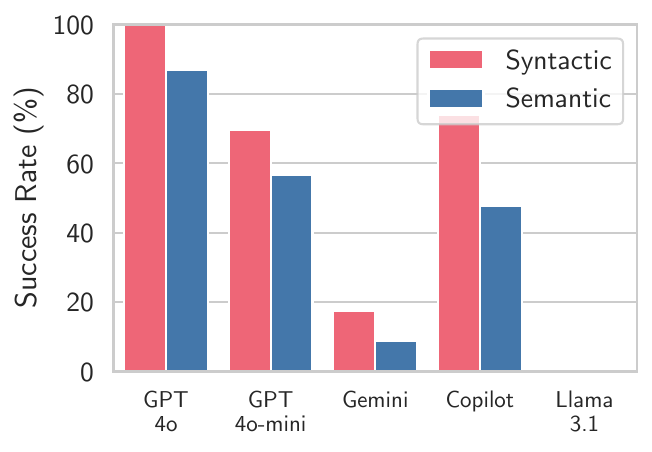}
        \caption{Easy}
        \label{fig:en-easy}
    \end{subfigure}%
    \hfill
    \begin{subfigure}{0.49\linewidth}
        \centering
        \includegraphics[width=\textwidth]{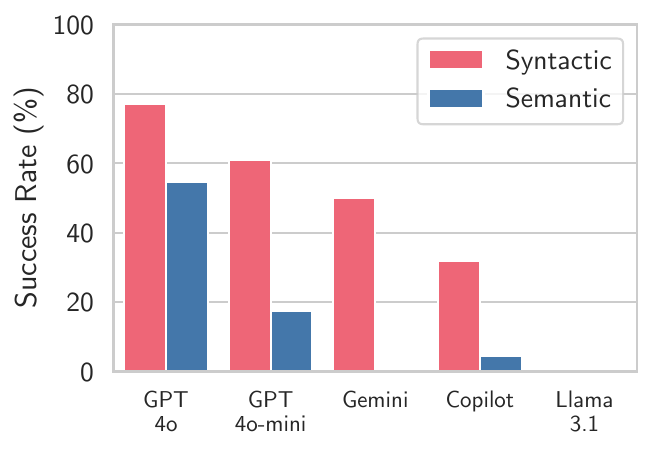}
        \caption{Medium}
        \label{fig:en-medium}
    \end{subfigure}%
    \vfill
    \begin{subfigure}{0.49\linewidth}
        \centering
        \includegraphics[width=\textwidth]{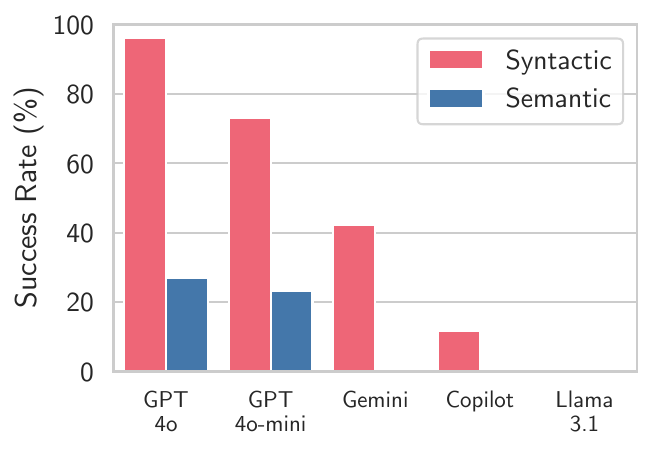}
        \caption{Advanced}
        \label{fig:en-advanced}
    \end{subfigure}%
    \hfill
    \begin{subfigure}{0.49\linewidth}
        \centering
        \includegraphics[width=\textwidth]{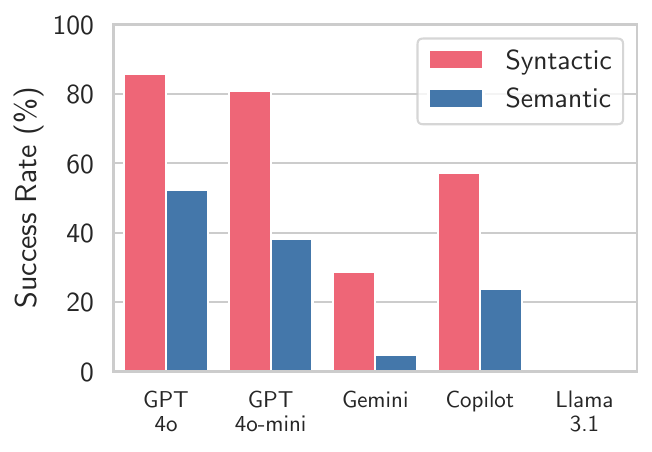}
        \caption{Complex}
        \label{fig:en-complex}
    \end{subfigure}%
    \caption{
        Success rates with English prompts.
        \emph{GPT-4o} still outperforms other models with English prompts in both syntactic and semantic success.
    }
    \label{fig:en-semantic-syntactic}
\end{figure*}

\begin{figure*}
    \centering
    \begin{subfigure}{0.49\linewidth}
        \includegraphics[width=\textwidth]{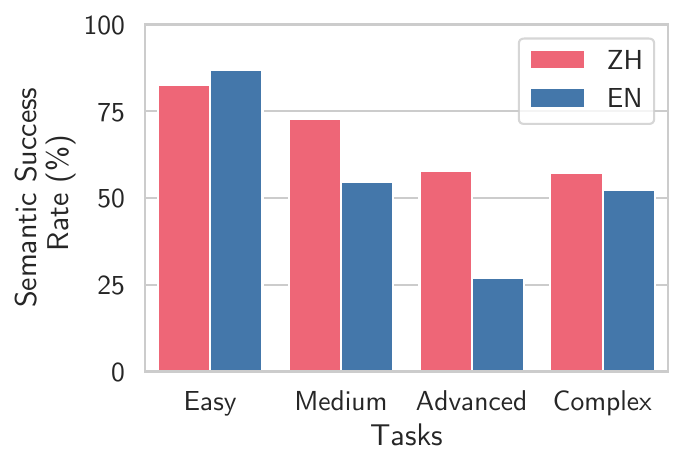}
        \caption{GPT-4o}
        \label{fig:result:language-semantic:gpt-4o}
    \end{subfigure}
    \hfill
    \begin{subfigure}{0.49\linewidth}
        \includegraphics[width=\textwidth]{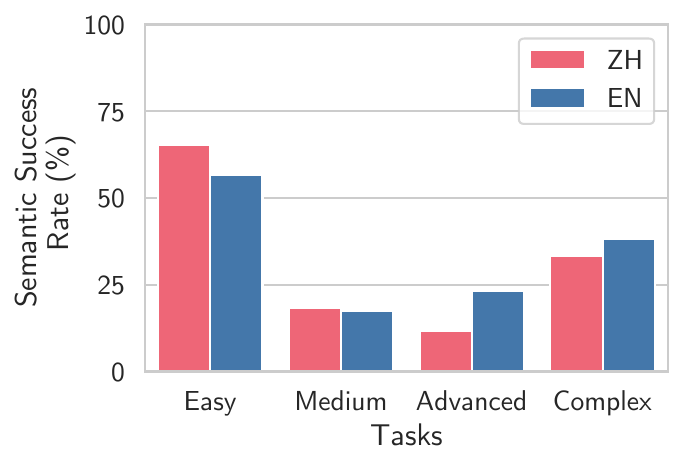}
        \caption{GPT-4o-mini}
        \label{fig:result:language-semantic:gpt-4o-mini}
    \end{subfigure}
    \hfill
    \begin{subfigure}{0.49\linewidth}
        \includegraphics[width=\textwidth]{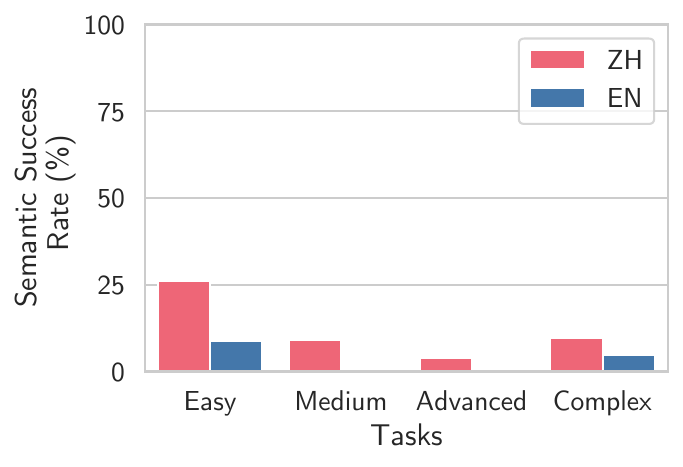}
        \caption{Gemini}
        \label{fig:result:language-semantic:gemini}
    \end{subfigure}
    \hfill
    \begin{subfigure}{0.49\linewidth}
        \includegraphics[width=\textwidth]{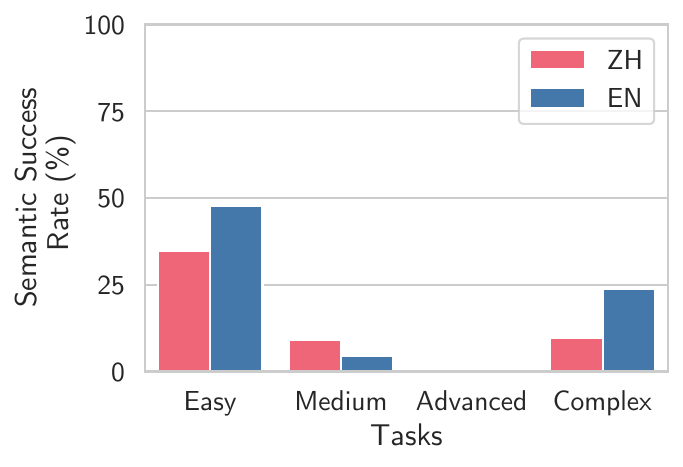}
        \caption{Copilot}
        \label{fig:result:language-semantic:copilot}
    \end{subfigure}

    \caption{Semantic success rate with prompts in Chinese and English.
        Translating user requirements from Chinese (ZH) to English (EN) does not significantly affect the semantic success rate.
        For results in both languages, \emph{OpenAI} models show a clear advantage.
    }
    \label{fig:result:language-semantic}
\end{figure*}

\begin{figure*}
    \centering
    \begin{subfigure}{0.49\linewidth}
        \includegraphics[width=\textwidth]{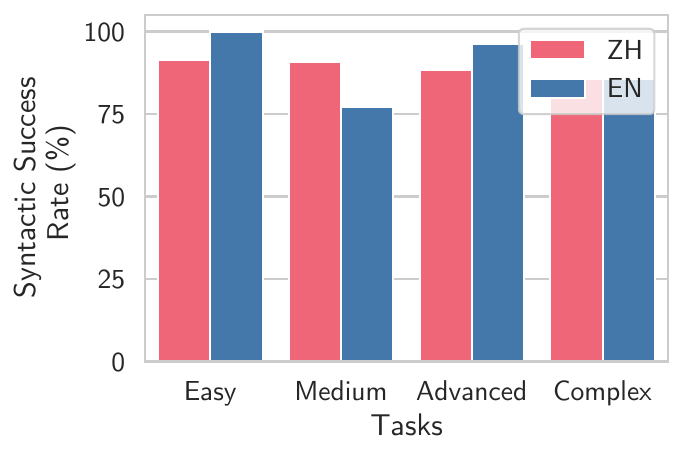}
        \caption{GPT-4o}
        \label{fig:result:language-syntactic:gpt-4o}
    \end{subfigure}
    \hfill
    \begin{subfigure}{0.49\linewidth}
        \includegraphics[width=\textwidth]{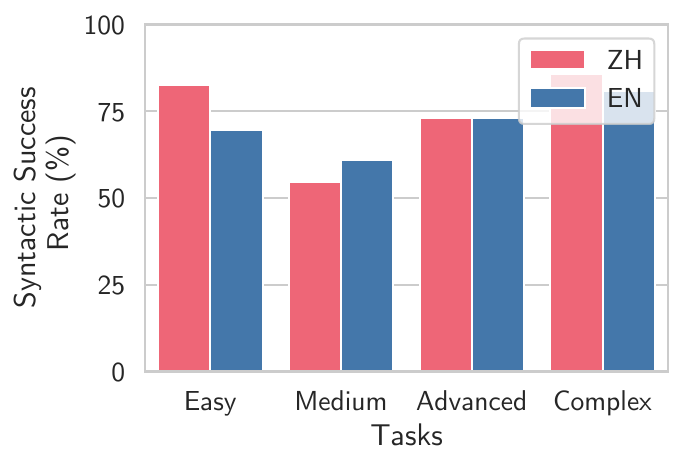}
        \caption{GPT-4o-mini}
        \label{fig:result:language-syntactic:gpt-4o-mini}
    \end{subfigure}
    \hfill
    \begin{subfigure}{0.49\linewidth}
        \includegraphics[width=\textwidth]{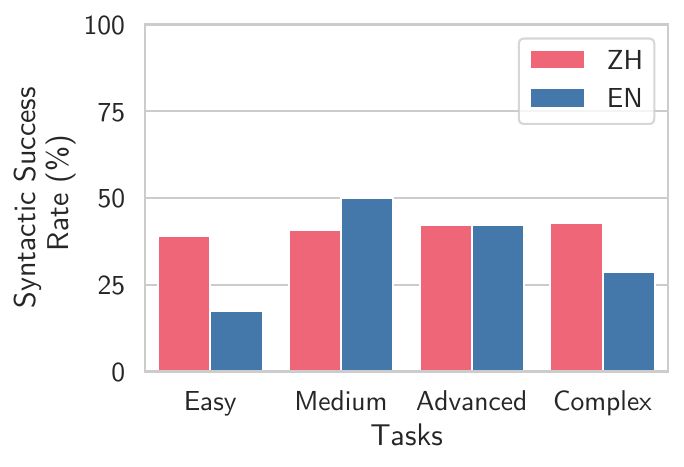}
        \caption{Gemini}
        \label{fig:result:language-syntactic:gemini}
    \end{subfigure}
    \hfill
    \begin{subfigure}{0.49\linewidth}
        \includegraphics[width=\textwidth]{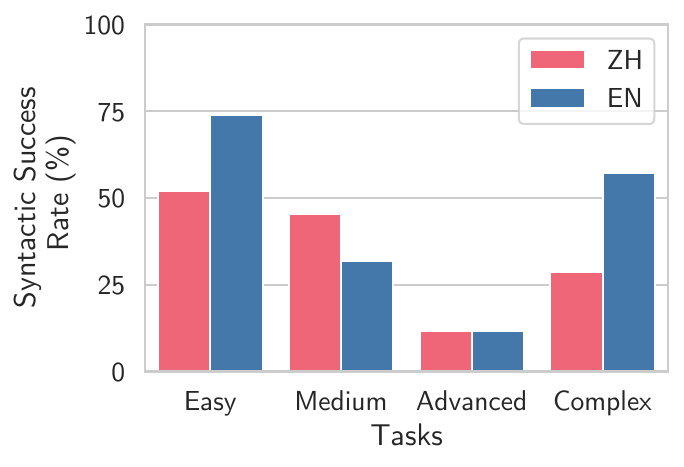}
        \caption{Copilot}
        \label{fig:result:language-syntactic:copilot}
    \end{subfigure}

    \caption{Syntactic success rate with prompts in Chinese (ZH) and English (EN).
        The language of user prompts has little impact on the syntactic success rate.
        \emph{GPT-4o} and \emph{GPT-4o-mini} maintain a high syntactic success rate with English prompts.
    }
    \label{fig:result:language-syntactic}
\end{figure*}

To investigate the impact of prompt language to LLM performance, we use the translated English prompt from dataset and evaluate the performance among the five LLMs.
We show the syntactic and semantic success rate of English prompt results in \cref{fig:en-semantic-syntactic}, then show the comparison to original results in
\cref{fig:result:language-semantic} and \cref{fig:result:language-syntactic}.

\emph{GPT-4o} and \emph{GPT-4o-mini}, i.e., OpenAI models, maintain relatively high average syntactic success rates, i.e. 89.78\% and 71.12\%, respectively, while the other LLMs remain below 50\%.
\emph{LLaMA} still fails to generate functions for all the prompts, resulting in 0\% syntactic and semantic success rates.
For \emph{GPT-4o}, the syntactic success rate increases slightly by 0.69 percentage points, while the semantic success rate decreases by 12.34 percentage points compared to the original prompts.
English prompts improve performance only on easy tasks but negatively impact more complex ones.
The average syntactic and semantic success rates of \emph{GPT-4o-mini} almost remain unchanged, with a 1.68 percentage point increase in semantic success rate and a 2.87 percentage point decrease in syntactic success rate.
For tasks in different complexity levels, it shows an opposite trend to \emph{GPT-4o}, where it drops for easy and medium tasks but improves for advanced and complex ones.
The translation step helps \emph{GPT-4o-mini} to better understand the user requirements with advanced and complex tasks, while losing nuances of the original user requirements leads to a performance drop in the easy and medium tasks.
\emph{Gemini} experiences a decrease in both syntactic and semantic success rates by 8.77 percentage points and 6.73 percentage points, respectively.
\emph{Copilot}, in contrast, demonstrates an overall improvement of 9.17 percentage points in syntactic success and 5.7 percentage points in semantic success.
Specifically, with English prompts, \emph{Copilot} achieves a comparable performance to \emph{GPT-4o-mini} with the easy task.

\subsection{Community Background}
\label{sec:results:community}

\begin{figure*}
    \centering
    \begin{subfigure}{0.49\linewidth}
        \centering
        \includegraphics[width=\textwidth]{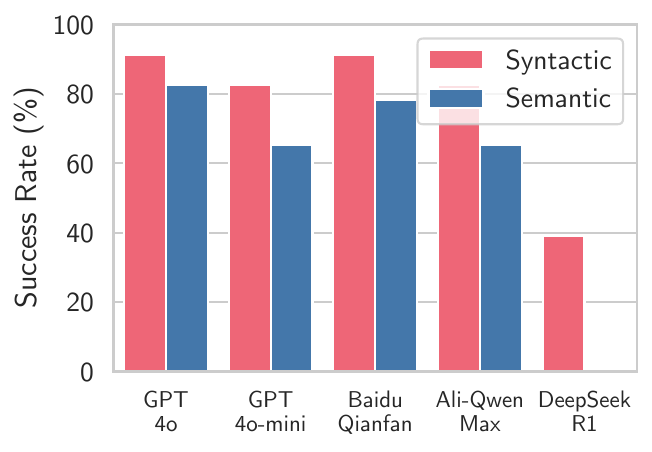}
        \caption{Easy}
        \label{fig:mandarin-easy}
    \end{subfigure}%
    \hfill
    \begin{subfigure}{0.49\linewidth}
        \centering
        \includegraphics[width=\textwidth]{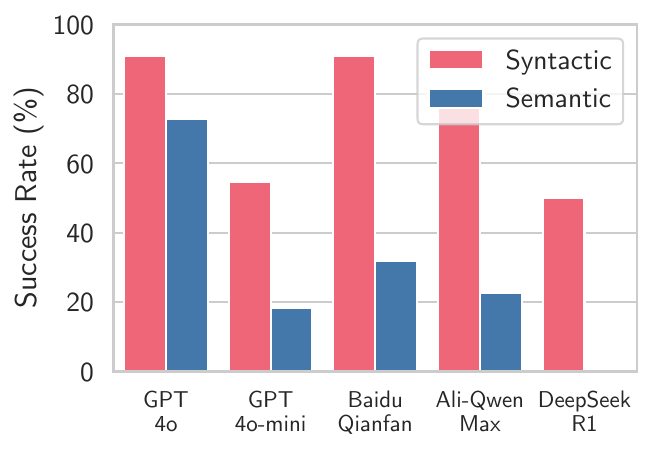}
        \caption{Medium}
        \label{fig:mandarin-medium}
    \end{subfigure}%
    \vfill
    \begin{subfigure}{0.49\linewidth}
        \centering
        \includegraphics[width=\textwidth]{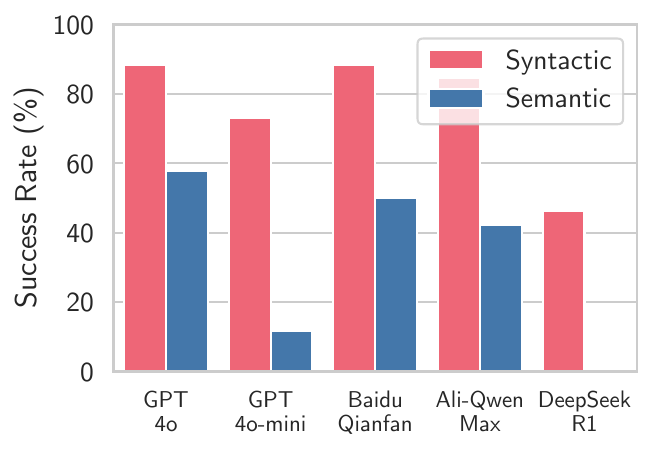}
        \caption{Advanced}
        \label{fig:mandarin-advanced}
    \end{subfigure}%
    \hfill
    \begin{subfigure}{0.49\linewidth}
        \centering
        \includegraphics[width=\textwidth]{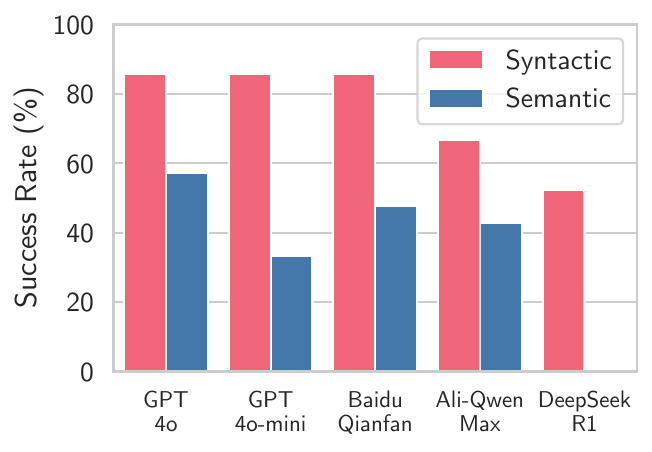}
        \caption{Complex}
        \label{fig:mandarin-complex}
    \end{subfigure}%
    \caption{Comparison of OpenAI models with Chinese  LLMs: \emph{GPT-4o} still outperforms other models in both syntactic and semantic accuracy, but the Chinese models show a clear advantage to \emph{GPT-4o-mini}.
        \emph{Baidu-Qianfan} can reach a similar semantic performance in easy tasks and syntactic performance with \emph{GPT-4o} for all difficulty levels.
    }
    \label{fig:mandarin-semantic-syntactic}
\end{figure*}

\begin{figure*}
    \centering
    \begin{subfigure}{0.49\linewidth}
        \includegraphics[width=\textwidth]{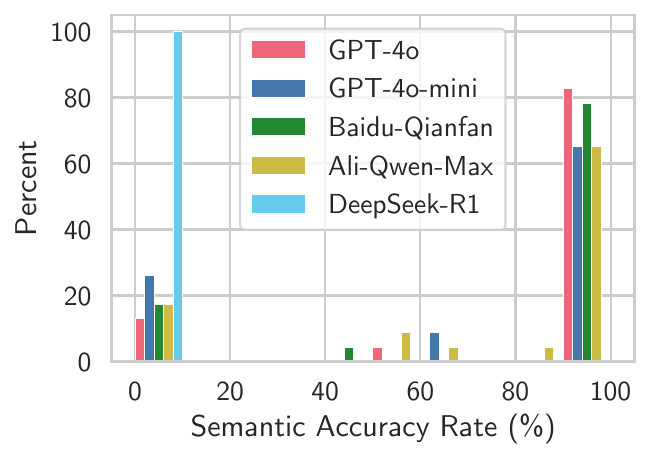}
        \caption{Easy}
        \label{fig:result:mandarin-count-easy}
    \end{subfigure}
    \hfill
    \begin{subfigure}{0.49\linewidth}
        \includegraphics[width=\textwidth]{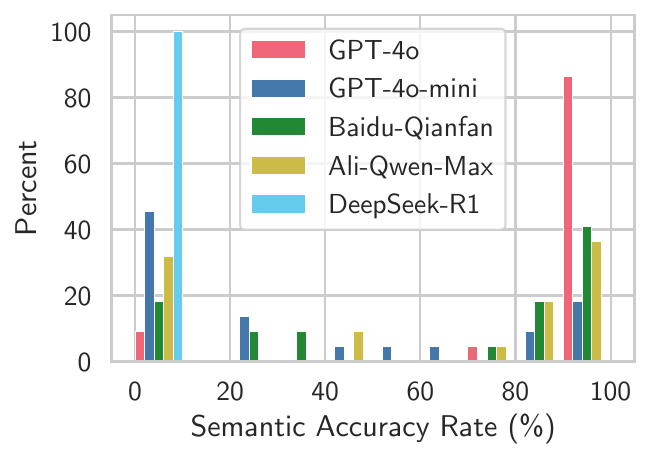}
        \caption{Medium}
        \label{fig:result:mandarin-count-medium}
    \end{subfigure}
    \hfill
    \begin{subfigure}{0.49\linewidth}
        \includegraphics[width=\textwidth]{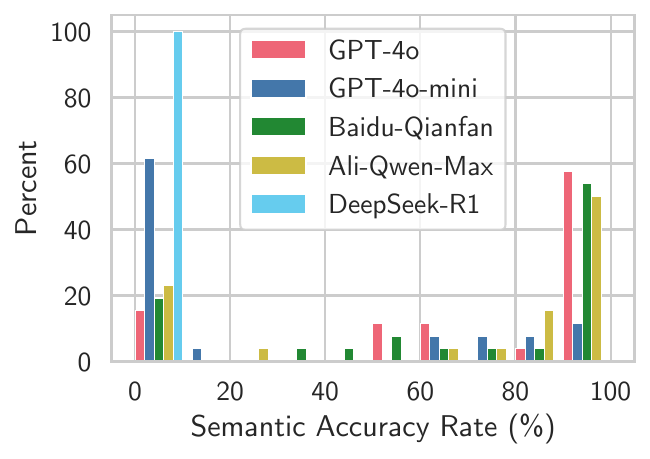}
        \caption{Advanced}
        \label{fig:result:mandarin-count-advanced}
    \end{subfigure}
    \hfill
    \begin{subfigure}{0.49\linewidth}
        \includegraphics[width=\textwidth]{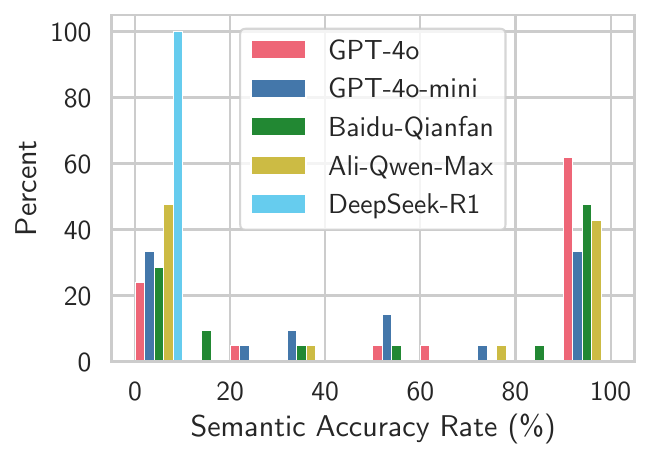}
        \caption{Complex}
        \label{fig:result:mandarin-count-complex}
    \end{subfigure}

    \caption{Semantic Accuracy Comparison with Chinese Models:
        \emph{GPT-4o} also outperforms the Chinese models in semantic accuracy across tasks of varying difficulty, but the differences are less pronounced.
        The two Chinese models showcase a similar performance of semantic accuracy where \emph{Baidu-Qianfan} is slightly better.
        With the task complexity increase, the semantic accuracy distribution of results becomes more dispersed for all models.
    }
    \label{fig:result:mandarin-semantic-count}
\end{figure*}

Considering LLMs trained with language-specific data may have a better understanding of the user prompts in the same language, we evaluate the performance of three Chinese LLMs, i.e., \emph{Baidu-Qianfan}, \emph{Ali-Qwen-Max} and \emph{DeepSeek-r1:7b}, with original Chinese user requirements.
We compare their performance with the \emph{OpenAI} models, which performed best in the initial experiments and show the results in \cref{fig:mandarin-semantic-syntactic} and \cref{fig:result:mandarin-semantic-count}.

\emph{Baidu-Qianfan} shows a same syntactic performance to \emph{GPT-4o} with an 89.10\% syntactic success rate, followed by \emph{Ali-Qwen-Max}, \emph{GPT-4o-mini} and \emph{DeepSeek} with 77.79\%, 73.99\% and 46.92\%, respectively.
In terms of the average semantic success rate of all tasks, \emph{GPT-4o} maintains its lead with 67.54\%. 
\emph{Baidu-Qianfan}, \emph{Ali-Qwen-Max}, and \emph{GPT-4o-mini} follow with success rates of 51.93\%, 43.28\%, and 32.07\%, respectively.
\emph{DeepSeek} entirely fails to generate function that meets user request, lead to a 0\% semantic success rate.
Notably, \emph{Baidu-Qianfan} can achieve a similar semantic success rate (78.26\%) to \emph{GPT-4o} (82.61\%) in the easy task but not for more complex ones.
\emph{Ali-Qwen-Max} has a comparable performance to \emph{GPT-4o-mini} in the easy task, and shows a clear advantage in the tasks with higher complexity.
Excluding \emph{DeepSeek} which has a 0\% semantic success rate, \emph{GPT-4o-mini} has the worst performance among this comparison, which can be caused by both the model capability and understanding of prompt language.

\subsection{Zero-shot vs. Few-shot Comparison}
\label{sec:results:zero_shot_few_shot}

\begin{figure*}
    \centering
    \begin{subfigure}{0.49\linewidth}
        \includegraphics[width=\textwidth]{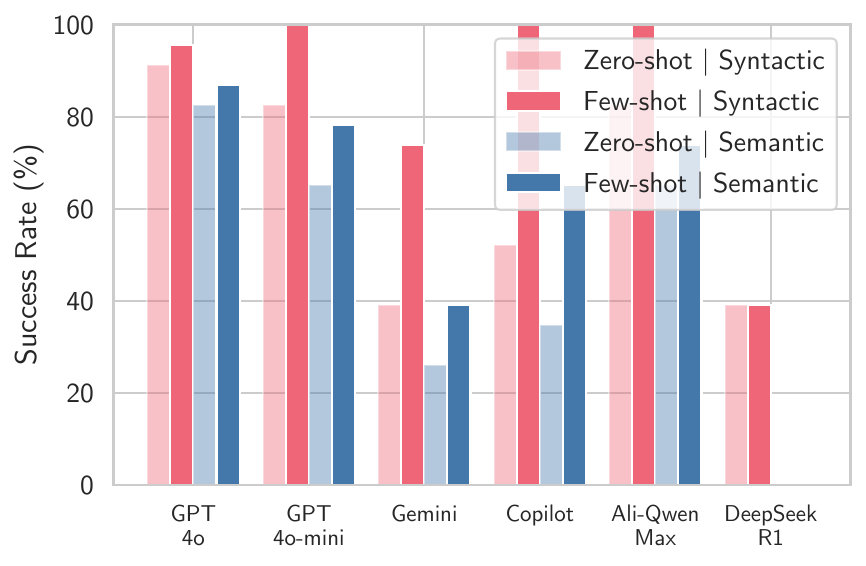}
        \caption{Easy}
        \label{fig:result:shot-compare-easy}
    \end{subfigure}
    \hfill
    \begin{subfigure}{0.49\linewidth}
        \includegraphics[width=\textwidth]{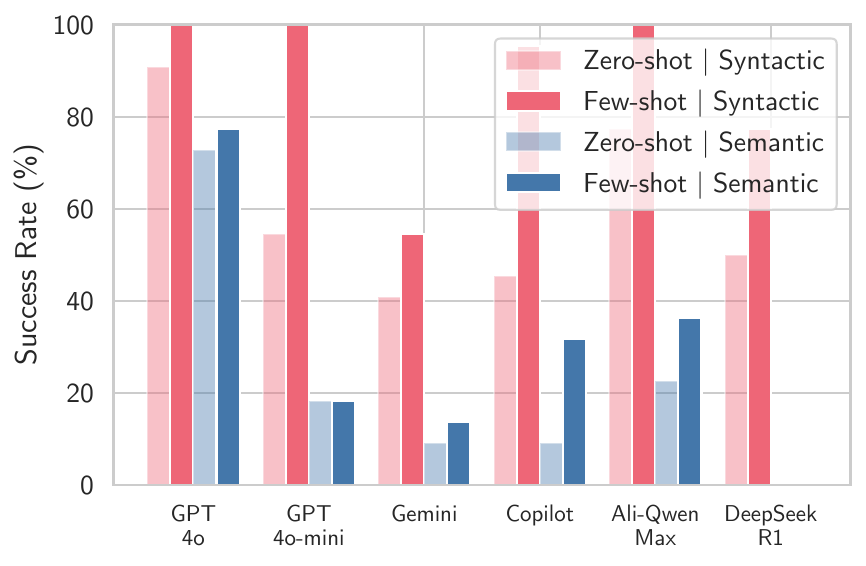}
        \caption{Medium}
        \label{fig:result:shot-compare-medium}
    \end{subfigure}
    \hfill
    \begin{subfigure}{0.49\linewidth}
        \includegraphics[width=\textwidth]{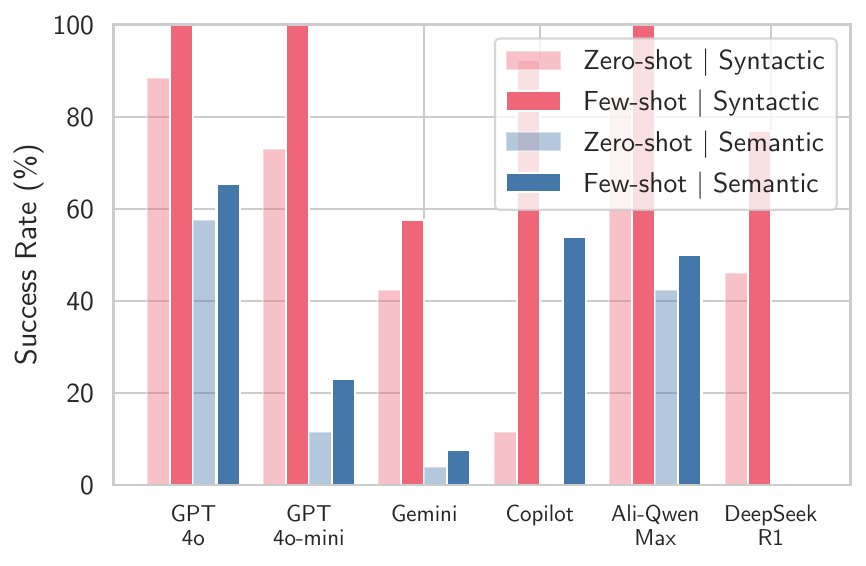}
        \caption{Advanced}
        \label{fig:result:shot-compare-advanced}
    \end{subfigure}
    \hfill
    \begin{subfigure}{0.49\linewidth}
        \includegraphics[width=\textwidth]{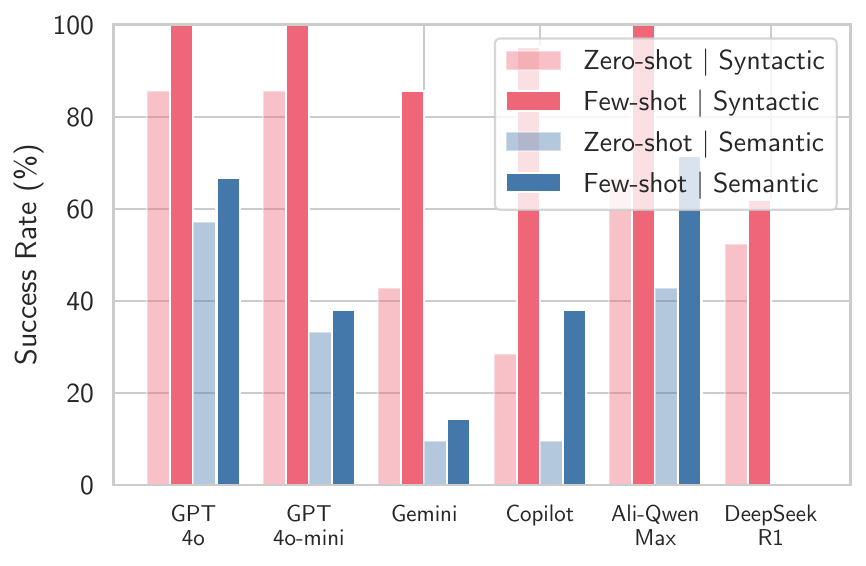}
        \caption{Complex}
        \label{fig:result:shot-compare-complex}
    \end{subfigure}

    \caption{Syntactic and Semantic Success Rate after Few-Shot Experiment with all models.
    We show few-shot and zero-shot results using high and low opacity bars, respectively.
    For all models, giving error messages as feedback can improve both the syntactic and semantic success rate. }
    \label{fig:result:shot-compare-semantic-syntactic}
\end{figure*}

\begin{figure*}
    \centering
    \begin{subfigure}{0.49\linewidth}
        \includegraphics[width=\textwidth]{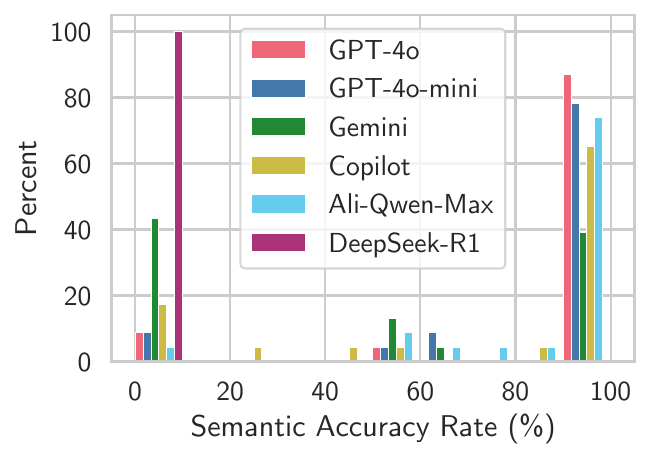}
        \caption{Easy}
        \label{fig:result:few-shot-all-count-easy}
    \end{subfigure}
    \hfill
    \begin{subfigure}{0.49\linewidth}
        \includegraphics[width=\textwidth]{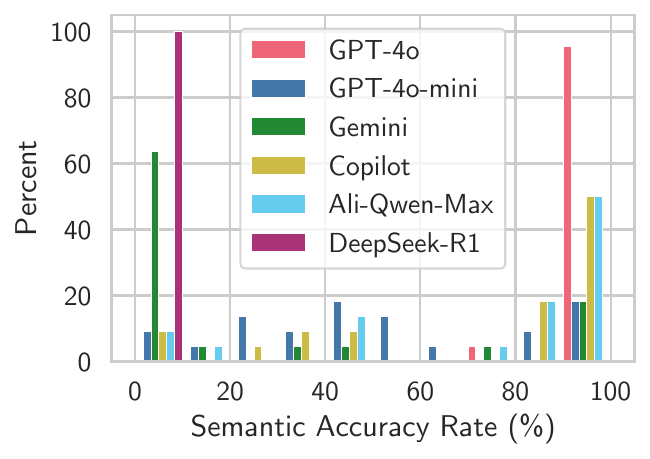}
        \caption{Medium}
        \label{fig:result:few-shot-all-count-medium}
    \end{subfigure}
    \hfill
    \begin{subfigure}{0.49\linewidth}
        \includegraphics[width=\textwidth]{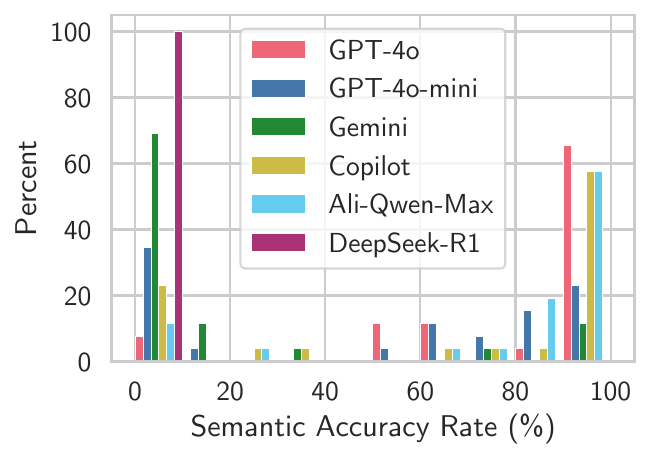}
        \caption{Advanced}
        \label{fig:result:few-shot-all-count-advanced}
    \end{subfigure}
    \hfill
    \begin{subfigure}{0.49\linewidth}
        \includegraphics[width=\textwidth]{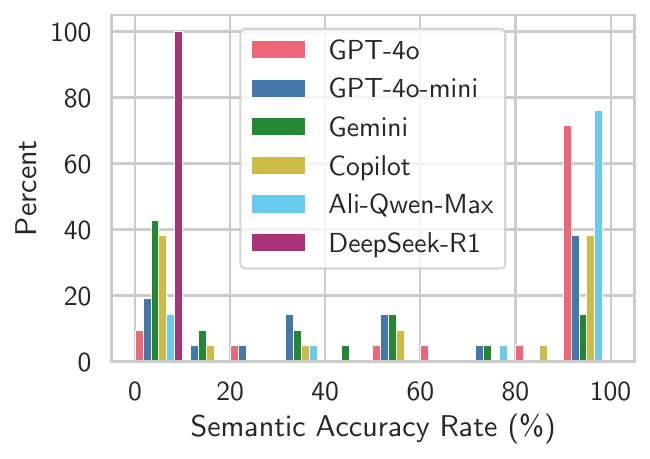}
        \caption{Complex}
        \label{fig:result:few-shot-all-count-complex}
    \end{subfigure}

    \caption{Semantic Accuracy of Few-Shot Experiment with Prompt in Chinese among all models.
    The semantic accuracy shows a similar bimodal distribution as the zero-shot setting results, while \emph{GPT-4o} and \emph{Ali-Qwen-Max} tends to be more concentrated around 100\% semantic accuracy among all tasks.
    }
    \label{fig:result:few-shot-zh-all-semantic-count}
\end{figure*}

To better understand the effect of iterative feedback on code generation performance, we compare the zero-shot and few-shot prompting strategies across multiple LLMs. 
This evaluation aims to quantify how compiler-level feedback, specifically, syntactic error messages, can help LLMs refine their outputs.
In the few-shot experiments, we focus exclusively on syntactic errors, as these can be reliably and objectively detected by compilers.
In contrast, semantic errors are inherently more difficult to define and evaluate, particularly in the NCDPs context, where non-technical users may express requirements in diverse and subjective ways.
We evaluate the same set of LLMs as in previous experiments, including \emph{GPT-4o}, \emph{GPT-4o-mini}, \emph{Gemini-1.5-flash}, \emph{Copilot}, \emph{Ali-Qwen-Max}, and \emph{DeepSeek-r1:7b}.
We exclude \emph{Baidu-Qianfan} and \emph{LLaMA} from the few-shot experiment, as incorporating error messages as feedback to LLM would exceed the input token limit of \emph{Baidu-Qianfan}, and \emph{LLaMA} exhibits an all-zero performance in the zero-shot setting.
Also, we use the English prompts in the few-shot experiment to showcase the input language influence. 

\subsubsection{Few-Shot: Model Selection}
Providing error messages as feedback consistently enhances syntactic success across all models, with some achieving 100\% syntactic success.
This syntactic correction also leads to improvements in semantic success.
We show the syntactic and semantic success rate of few-shot results in \cref{fig:result:shot-compare-semantic-syntactic}, and present the semantic accuracy distribution in \cref{fig:result:few-shot-zh-all-semantic-count}.

The overall performance ranking in the few-shot setting differs slightly from the zero-shot setting. 
While \emph{GPT-4o} remains the top model, \emph{GPT-4o-mini}, which ranked second in the zero-shot experiments, drops to third place, overtaken by \emph{Copilot}.
In detail, \emph{GPT-4o} and \emph{GPT-4o-mini}, i.e., the top-performing models in the zero-shot setting, continue to show strong performance in the few-shot settings.
Specifically, \emph{GPT-4o-mini} achieves a 100\% syntactic success rate for all tasks, while \emph{GPT-4o} fails to resolve one syntactic error, resulting in a 95.65\% syntactic success rate, all other tasks yield 100\% syntactic success.
Regarding semantic success rate, \emph{GPT-4o} achieves an average of 76.57\% across all tasks, while \emph{GPT-4o-mini} has a 39.41\% semantic success rate, comparing to the 67.54\% and 32.07\% in the zero-shot setting.
Additionally, we measure the semantic accuracy, representing how close the model outputs are to semantic success.
\emph{GPT-4o} achieves an average of 87.65\% across all tasks, while \emph{GPT-4o-mini} has a 61.42\% semantic accuracy, comparing to the 79.20\% and 45.97\% in the zero-shot setting.

In the few-shot setting, \emph{Copilot} shows substantial improvements in both syntactic and semantic success rate, achieving an average syntactic success rate of 95.75\% and an average semantic success rate of 47.25\% across all tasks, comparing to 34.43\% and 13.35\%, respectively, in the zero-shot setting.
\emph{Copilot} does not achieve a 100\% syntactic success rate as \emph{GPT-4o-mini}, however, it surpasses it in terms of average semantic success rate.
In particular, \emph{Copilot} outperforms \emph{GPT-4o-mini} on all but the easy task when measuring semantic success.
\emph{Copilot} tends to first generate a skeleton or pseudocode of the user required function in the first round, and fill the function logic in the following iterations, i.e., it acts more like a human developer.
It demonstrates a 66.49\% semantic accuracy rate in the few-shot setting, which is an improvement over the zero-shot setting at 19.78\%.

\emph{Gemini} showcases a syntactic improvement compared to the zero-shot setting, however, this enhancement has limited impact on its semantic performance.
Except for the easy task, where \emph{Gemini} achieves an average 39.13\% semantic success rate, other tasks are around 10\%.

\subsubsection{Few-Shot: Community Background}

We present the Chinese LLMs results alongside those of English LLMs results in \cref{fig:result:shot-compare-semantic-syntactic} and \cref{fig:result:few-shot-zh-all-semantic-count} because the performance gap among the LLMs is narrowing.
Considering all the LLMs evaluated in the few-shot setting, the performance ranking is \emph{GPT-4o}, \emph{Ali-Qwen-Max}, \emph{Copilot}, \emph{GPT-4o-mini}, \emph{Gemini}, and \emph{DeepSeek R1}.

In the few-shot setting, \emph{Ali-Qwen-Max} continues to demonstrate strong performance and significant improvements.
It achieves a syntactic success rate of 100\% across all tasks, and an average semantic success rate of 57.93\%, up from 77.79 and 43.28\%, respectively, in the zero-shot setting.
The semantic accuracy of \emph{Ali-Qwen-Max} also improves to an average of 81.02\%, compared to 63.54\% in the zero-shot setting, narrowing the gap with \emph{GPT-4o} at 87.65\%.
In the complex task, \emph{Ali-Qwen-Max} outperforms \emph{GPT-4o} in the semantic success rate, achieving 71.43\% compared to \emph{GPT-4o}'s 66.67\%.
While it can be partially attributed to a higher number of syntactic error cases, \emph{Ali-Qwen-Max} still shows a strong performance in resolving syntactic errors and improving semantic success performance.

\emph{DeepSeek R1} also shows a syntactic improvement, increasing from 46.92\% to 63.81\%. 
However, it still fails to achieve any semantic success or semantic accuracy among all tasks.

\subsubsection{Few-Shot: Prompt Language}

\begin{figure*}
    \centering
    \begin{subfigure}{0.49\linewidth}
        \includegraphics[width=\textwidth]{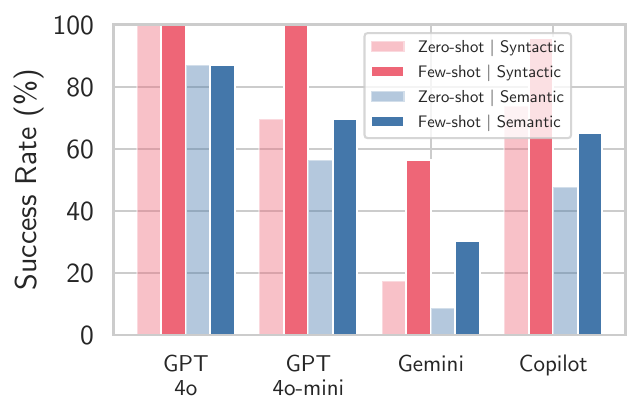}
        \caption{Easy}
        \label{fig:result:few-shot-en-barplot-easy}
    \end{subfigure}
    \hfill
    \begin{subfigure}{0.49\linewidth}
        \includegraphics[width=\textwidth]{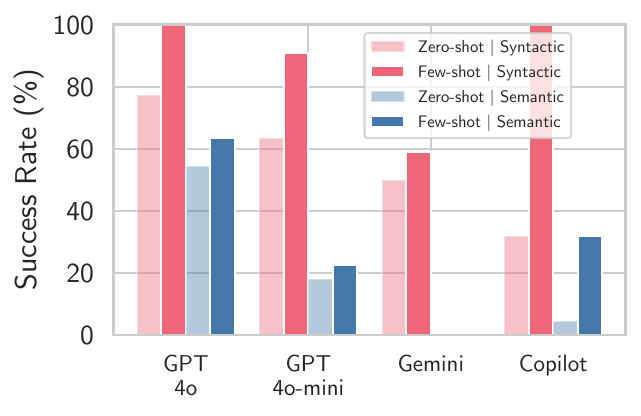}
        \caption{Medium}
        \label{fig:result:few-shot-en-barplot-medium}
    \end{subfigure}
    \hfill
    \begin{subfigure}{0.49\linewidth}
        \includegraphics[width=\textwidth]{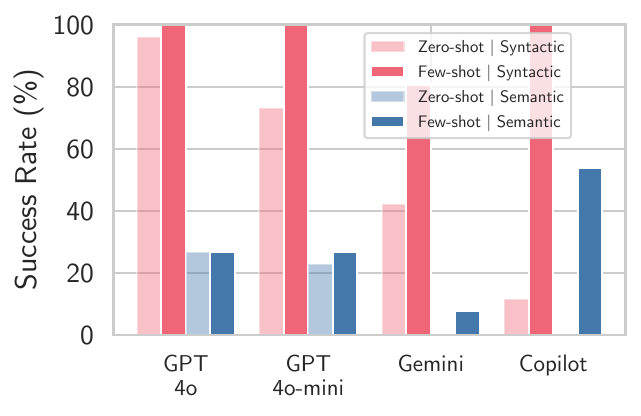}
        \caption{Advanced}
        \label{fig:result:few-shot-en-barplot-advanced}
    \end{subfigure}
    \hfill
    \begin{subfigure}{0.49\linewidth}
        \includegraphics[width=\textwidth]{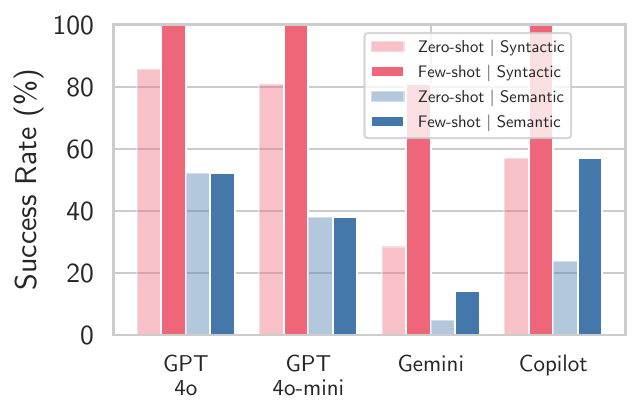}
        \caption{Complex}
        \label{fig:result:few-shot-en-barplot-complex}
    \end{subfigure}

    \caption{
    Syntactic and Semantic Success Rate after Few-Shot Experiment with English prompts.
    Similar with using Chinese prompts, the few-shot experiments with English prompts also improves in syntactic and semantic success rate for all models.
    }
    \label{fig:result:few-shot-en-barplot}
\end{figure*}

\begin{figure*}
    \centering
    \begin{subfigure}{0.49\linewidth}
        \includegraphics[width=\textwidth]{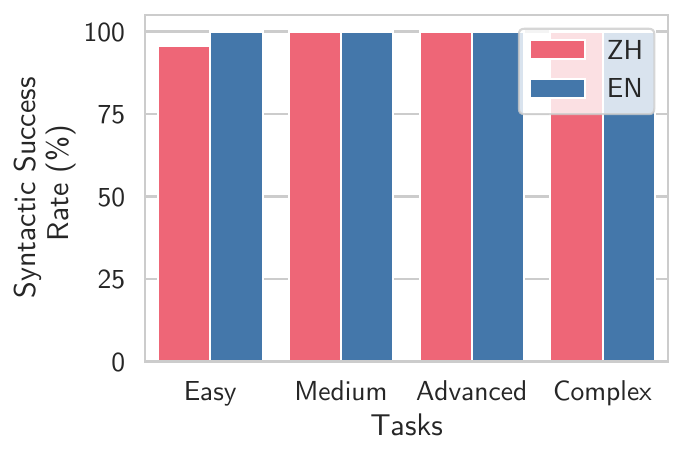}
        \caption{GPT-4o}
        \label{fig:result:few-shot-language-gpt4o}
    \end{subfigure}
    \hfill
    \begin{subfigure}{0.49\linewidth}
        \includegraphics[width=\textwidth]{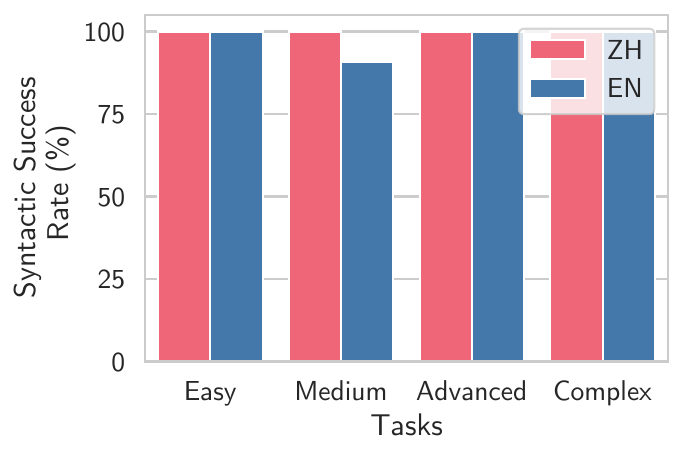}
        \caption{GPT-4o-mini}
        \label{fig:result:few-shot-language-gpt4o-mini}
    \end{subfigure}
    \hfill
    \begin{subfigure}{0.49\linewidth}
        \includegraphics[width=\textwidth]{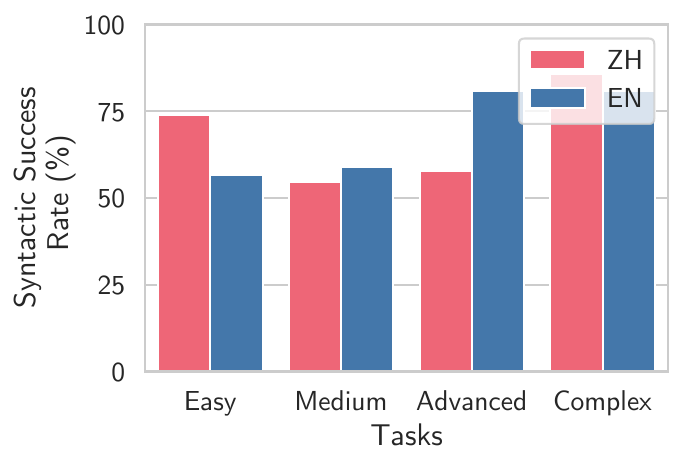}
        \caption{Gemini}
        \label{fig:result:few-shot-language-gemini}
    \end{subfigure}
    \hfill
    \begin{subfigure}{0.49\linewidth}
        \includegraphics[width=\textwidth]{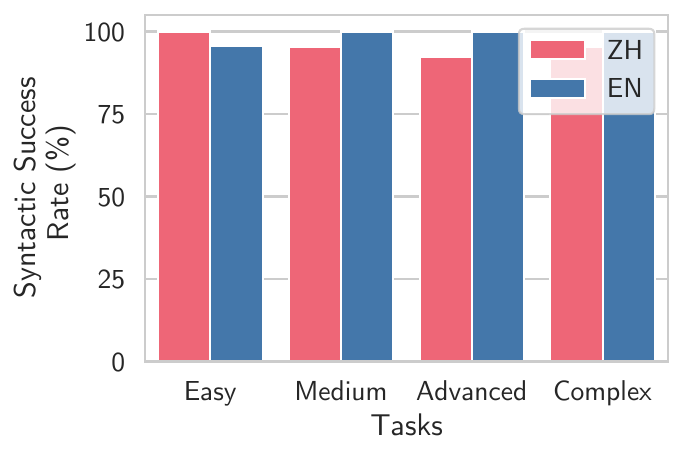}
        \caption{Copilot}
        \label{fig:result:few-shot-language-copilot}
    \end{subfigure}

    \caption{Syntactic success rate comparison with prompts in Chinese and English under the few-shot setting.
    We denote Chinese and English prompts as ZH and EN in the figure, respectively.
    The language choice of the prompt has little impact on the syntactic success rate.}
    \label{fig:result:few-shot-language-syntactic}
\end{figure*}

\begin{figure*}
    \centering
    \begin{subfigure}{0.49\linewidth}
        \includegraphics[width=\textwidth]{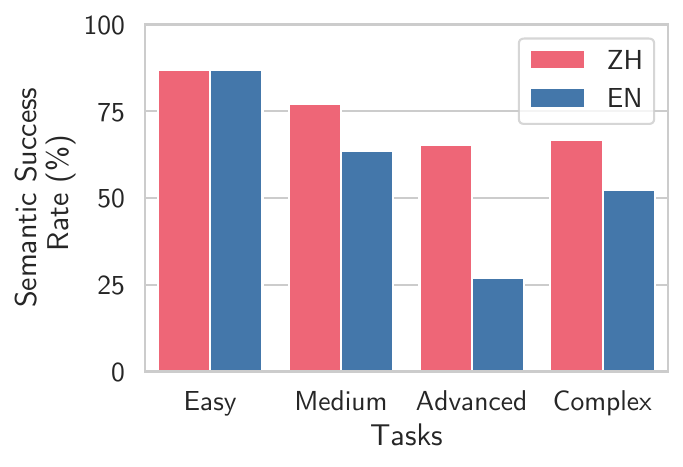}
        \caption{GPT-4o}
        \label{fig:result:few-shot-language-gpt4o}
    \end{subfigure}
    \hfill
    \begin{subfigure}{0.49\linewidth}
        \includegraphics[width=\textwidth]{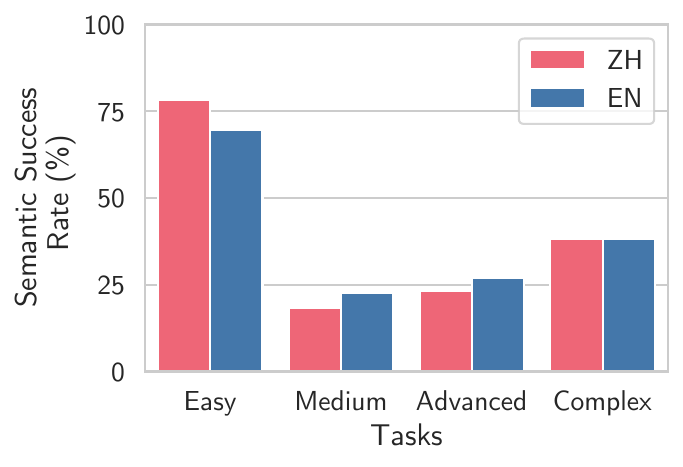}
        \caption{GPT-4o-mini}
        \label{fig:result:few-shot-language-gpt4o-mini}
    \end{subfigure}
    \hfill
    \begin{subfigure}{0.49\linewidth}
        \includegraphics[width=\textwidth]{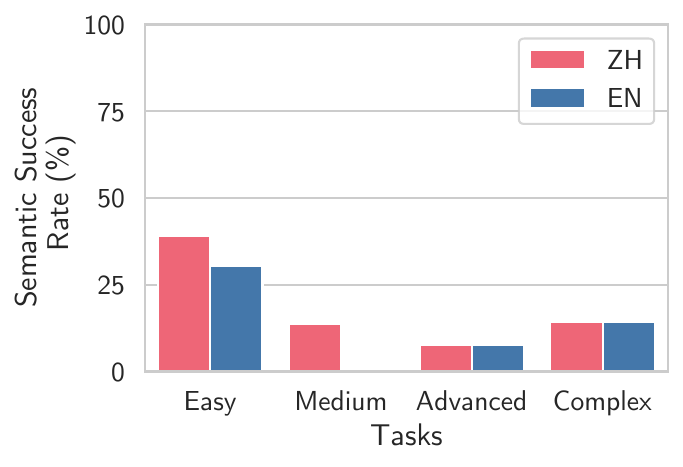}
        \caption{Gemini}
        \label{fig:result:few-shot-language-gemini}
    \end{subfigure}
    \hfill
    \begin{subfigure}{0.49\linewidth}
        \includegraphics[width=\textwidth]{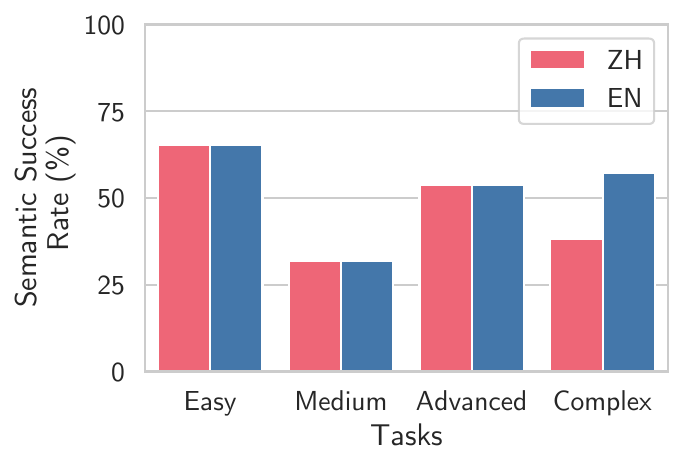}
        \caption{Copilot}
        \label{fig:result:few-shot-language-copilot}
    \end{subfigure}

    \caption{Semantic success rate comparison with Chinese and English prompts under the few-shot setting.
    We denote Chinese and English prompts as ZH and EN, respectively.
    \emph{GPT-4o} generally shows outstanding performance in semantic success rate under both Chinese and English prompts with few-shot setting.}
    \label{fig:result:few-shot-language-semantic}
\end{figure*}

Since prompt language can affect LLM performance, particularly when it aligns with the training data of LLMs, we also assess LLMs using English prompts under the few-shot setting.
We show the syntactic and semantic success rate of few-shot results in \cref{fig:result:few-shot-en-barplot}, and present the comparison with prompt in Chinese in \cref{fig:result:few-shot-language-syntactic} and \cref{fig:result:few-shot-language-semantic}.

The results indicate that all the models exhibit improvements in the syntactic success rate, but leading to different trend of improvement in the semantic success rate.
For the OpenAI models, i.e., \emph{GPT-4o} and \emph{GPT-4o-mini}, the few-shot experiment with English prompts improves the syntactic success rate but slightly improve the semantic success rate.
Specifically, for syntactic success rate, \emph{GPT-4o} improves from an average of 89.78\% to 100\%, and \emph{GPT-4o-mini} improves from 71.81\% to 97.73\%.
In terms of semantic success rate, \emph{GPT-4o} shows improvement only on the medium task, the others keep unchanged, leading to an average semantic success rate improvement of 2.28\%.
\emph{GPT-4o-mini} shows an average semantic improvement of 5.36\%, increasing from 33.97\% to 39.33\%, with no change only on the complex task.

\emph{Copilot}, on the other hand, demonstrates a significant improvement in both syntactic and semantic success rate among all tasks, with an average syntactic success rate of 98.91\% and an average semantic success rate of 52.01\%, comparing to those in the zero-shot setting of 43.60\% and 19.05\%, respectively.
Notably, \emph{Copilot} outperforms \emph{GPT-4o} for the advanced and complex tasks in the semantic success.
This is partly due to the low number of syntactic error cases of \emph{GPT-4o} in the few-shot setting, which limits opportunities for feedback-based improvement.
It demonstrates the coding ability of \emph{Copilot} with the pseudocode or code skeleton generated in the zero-shot experiment.
For the semantic accuracy, \emph{Copilot} achieves an average of 73.75\% across all tasks, a significant improvement compared over its zero-shot results at 25.45\%.
While \emph{Gemini} shows notable improvement in syntactic success rate, from 34.58\% to 69.33\%, its improved average semantic success rate remains low at 13.10\%.
Overall, using English prompts can help \emph{GPT-4o-mini} and \emph{Copilot} to achieve performance improvement, particularly in semantic understanding and code generation quality under the few-shot setting.

Comparing the results with those using Chinese prompts, we find that for \emph{GPT-4o}, \emph{GPT-4o-mini} and \emph{Copilot}, iteratively incorporating error messages enables nearly 100\% syntactic success rates. 
Although \emph{Gemini} gains improvement, it still lags behind the others.
For \emph{GPT-4o} and \emph{GPT-4o-mini}, the prompt language have little impact on the syntactic success.
In contrast, both \emph{Copilot} and \emph{Gemini} syntactically perform better when using English prompts than Chinese prompts.
As for the semantic success rate, using the Chinese prompts which containing the original user description performs better with \emph{GPT-4o} and \emph{Gemini}.
Using English prompts helps \emph{Copilot} achieve a better semantic success due to the elimination of translation step during the generation process.
\emph{GPT-4o-mini} has a better semantic success rate with Chinese prompts for the easy task, as the task complexity increases, the English prompts may help to better understand the user intents.

\subsubsection{Iteration Required to Fix Syntactic Errors}
\begin{figure*}
    \centering
    \begin{subfigure}{0.49\linewidth}
        \includegraphics[width=\textwidth]{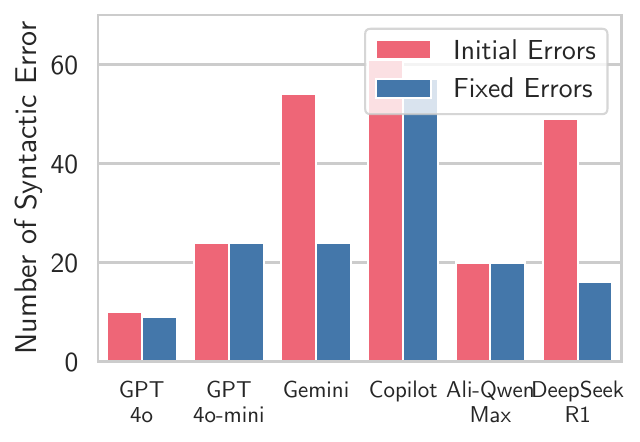}
        \caption{Experiment with Chinese Prompts}
        \label{fig:result:zh_error-fix-barplot}
    \end{subfigure}
    \hfill
    \begin{subfigure}{0.49\linewidth}
        \includegraphics[width=\textwidth]{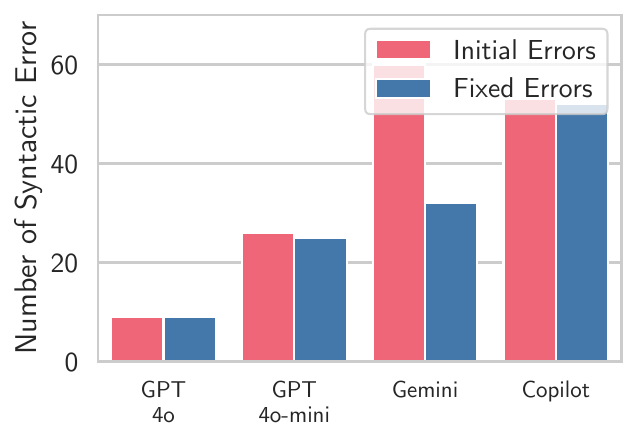}
        \caption{Experiment with English Prompts}
        \label{fig:result:en_error-fix-barplot}
    \end{subfigure}
    \caption{
        We show the amount of syntactic error cases of zero-shot experiment results as \emph{Initial Errors} for the few-shot setting.
        The \emph{Fixed Errors} bar shows the number of cases leads to a syntactic success within a maximum of 3 iterations.
    }
    \label{fig:result:few-shot_error-fix}
\end{figure*}

\begin{figure*}
    \centering
    \begin{subfigure}{0.49\linewidth}
        \includegraphics[width=\textwidth]{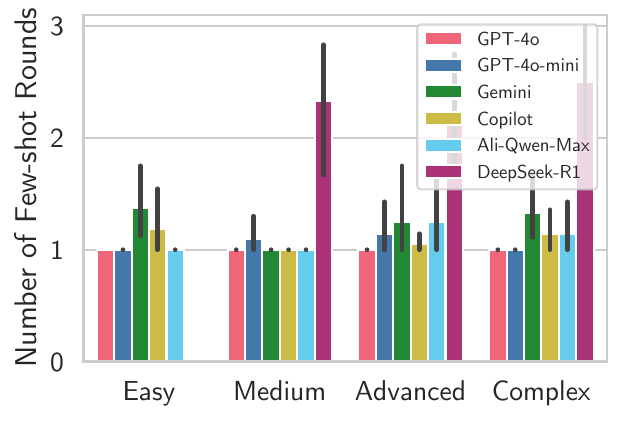}
        \caption{Experiment with Chinese Prompts}
        \label{fig:result:zh_iteration-barplot}
    \end{subfigure}
    \hfill
    \begin{subfigure}{0.49\linewidth}
        \includegraphics[width=\textwidth]{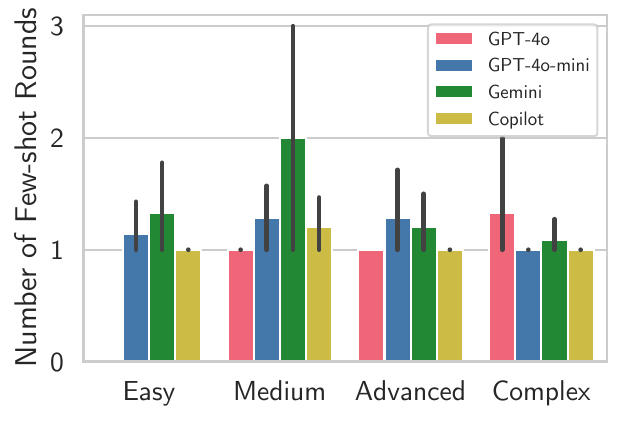}
        \caption{Experiment with English Prompts}
        \label{fig:result:en_iteration-barplot}
    \end{subfigure}

    \caption{
        We show the iterations needed to fix syntactic errors for the experiments with Chinese and English prompts among tasks in different levels of complexity.
        Zero represents initial syntactic error cases from zero-shot results.
    }
    \label{fig:result:few-shot-iteration}
\end{figure*}

In the few-shot experiments, we allow each LLM to iterate up to three times to resolve syntactic errors observed in the zero-shot results, and record the number of syntactic error cases that can be successfully corrected within these rounds.
We show the results in \cref{fig:result:few-shot_error-fix} and \cref{fig:result:few-shot-iteration}.

When using Chinese prompts, both \emph{GPT-4o-mini} and \emph{Ali-Qwen-Max} successfully resolved all syntactic errors within three iterations. 
\emph{GPT-4o} and \emph{Copilot} corrected 90.00\% and 93.44\% of syntactic errors, respectively.
Specifically, \emph{GPT-4o} failed to resolve 1 out of 10 error cases, while \emph{Copilot} failed in 4 out of 61 cases.
At the lower end of performance, \emph{Gemini} and \emph{DeepSeek R1} corrected only 44.44\% and 32.65\% of syntactic errors, respectively.
When using English prompts, \emph{GPT-4o} resolves all syntactic errors, followed by \emph{Copilot}, \emph{GPT-4o-mini} and \emph{Gemini} with 98.11\%, 96.15\%, and 53.33\% syntactic success rate, respectively.

While achieving high syntactic error fixing rates, the number of iterations required varies.  
Using Chinese prompts, \emph{GPT-4o} resolves all syntactic errors in the first iteration.
\emph{GPT-4o-mini} and \emph{Ali-Qwen-Max} can fix all syntactic errors within two rounds, i.e., around 90\% in the first iteration, and around 10\% in the second iteration.
With English prompts, \emph{GPT-4o} and \emph{GPT-4o-mini} fix errors within two iterations.
Specifically, \emph{GPT-4o} required a second round for 1 out of 9 cases.
\emph{Copilot} demonstrates strong performance by fixing nearly all syntactic errors in the first iteration, with only one case carried over to the second and one additional case requiring a third round out of 52 cases.

\subsection{Findings}
\label{sec:results:findings}
\paragraph{Feasibility of different LLMs}
Our evaluation across eight LLMs reveals a clear divergence in their ability to satisfy both syntactic and semantic requirements for NCDP.
According to the zero-shot experimental results, \emph{OpenAI} models, i.e., \emph{GPT-4o} and \emph{GPT-4o-mini}, and some Chinese LLMs, i.e., \emph{Baidu Qianfan} and \emph{Alibaba Qwen}, demonstrate strong performance in both syntactic and semantic accuracy, making them promising candidates for LLM-powered NCDPs.
All of them are designed as general purpose LLMs, without specialized optimization for tasks such as coding or reasoning.
In contrast, LLMs designed with auxiliary capabilities, e.g., \emph{Gemini}, \emph{Copilot}, and \emph{DeepSeek}, struggle to maintain both syntactic and semantic success.
These findings suggest that NCDP requires a balanced capability among different dimensions.
Specifically, \emph{GPT-4o} surpasses the Chinese LLMs in semantic accuracy while the Chinese LLMs exhibit a clear advantage over \emph{GPT-4o-mini} in both syntactic and semantic accuracy.
This suggests that the alignment of the language of LLM training data with the language of user requirements (e.g., Chinese) can enhance performance, however, the capability of LLM itself is also crucial.
Notably, a lightweight LLM, such as \emph{GPT-4o-mini}, performs well on simple tasks but struggles with complex ones, indicating that the size of LLM should be considered when selecting LLMs for real-world NCDPs.

\paragraph{Impact of Prompt Language}
The effect of prompt language (Chinese vs. translated English) reveals LLM-specific behaviors that point toward broader trends in language alignment.
Translating user requirements from Chinese to English has only a slight impact on the syntactic performance of \emph{OpenAI} models but affects semantic success more strongly, leading to a performance drop for \emph{GPT-4o} but an improvement for \emph{GPT-4o-mini}.
Although \emph{Gemini} and \emph{Copilot} have limited performance, translation improves both the syntactic and semantic accuracy of \emph{Copilot} while reducing these of \emph{Gemini}.
The variance in prompts has little effect on \emph{LLaMA}, which consistently fails to generate functions in both Chinese and English.
These mixed outcomes suggest that the impact of prompt language varies depending on the model's design or training focus. 
However, overall, prompt language alone does not affect the syntactic capabilities of high-performing LLMs, to which semantic factors remain more sensitive.

\paragraph{Alignment of LLM Performance with Task Complexity}
Interestingly, the performance of the LLMs does not entirely align with the predefined task complexity, i.e., LLMs occasionally perform better on the complex task than the medium and advanced tasks, except for the experiment setting of Chinese prompts used with \emph{GPT-4o}.
For results using translated English prompts, \emph{GPT-4o} also fails to maintain the alignment.
The possible reason for that is because both the medium and advanced tasks explicitly specify with three unrelated sub-tasks, each of which is relatively straightforward in terms of both user requirement description and the expected function logic.
Despite the straightforward requirements, the semantic success rate of these tasks drops significantly.
While the complex task involves a more complex functional logic and uncertainty and diversity of user preferences, it performs better.
Specifically, the Chinese LLMs exhibit similar performance alignment with what tasks specified, except for the medium task, where user requirements generally involve triggering more devices than the advanced task.
These findings suggest that while the task complexity affects the LLM performance, the presence of sub-tasks also plays a significant role.
For LLMs weak on natural language understanding, the task containing sub-tasks is more challenging than the complex one.
Additionally, a strong natural language understanding capability is essential for LLMs to generate correct results.

\paragraph{Reason for Performance Limitations}

According to the experimental results, the \emph{GPT-4o} and the Chinese LLMs demonstrate a strong performance in terms of syntactic and semantic accuracy.
In contrast, other LLMs struggle with semantic accuracy, though the underlying causes differ.
\begin{enumerate}
    \item \emph{LLaMA} either fails to generate a function or simply restructures and rewrites the reference code provided in the prompt.
    \item \emph{Gemini} generates an excessive amount of unnecessary code snippets irrelevant to user requirements.
          While some generated functions contain required functionalities, they often lack semantic accuracy.
    \item \emph{Copilot} primarily generates skeletal or pseudocode structures, offering little functional implementation.
    \item \emph{DeepSeek} mainly demonstrates a thinking process of both user requirements and reference code but produces incomplete lines of code.
\end{enumerate}
Overall, these limitations underscore the challenges that LLMs face in generating semantically correct and functional outputs.

\paragraph{Impact of Few-Shot Learning}
Few-shot prompting introduces measurable improvements in performance among all LLMs applied in the experiments, but its effectiveness varies across model types.
For high-performing LLMs such as \emph{GPT-4o}, which already achieve high syntactic success rates in zero-shot settings, while including a small amount of error cases, few-shot prompting further reduces residual errors in both syntactic and semantic perspectives.
LLMs exhibiting moderate semantic performance and including more error cases, such as \emph{GPT-4o-mini} and \emph{Ali-Qwen-Max}, benefit more from few-shot prompting, achieving a significant improvement in syntactic and semantic success rate.
LLMs with strong coding capabilities but weak zero-shot performance, i.e., \emph{Copilot}, also exhibit notable gains from few-shot iterations.
These LLMs tend to produce abstract pseudocode or skeletal structures in zero-shot scenarios, the few-shot feedback helps to infer functional logic and complete code structures, thus improving semantic accuracy.
In contrast, for LLMs with overall weak performance and limited code generation capabilities, i.e., \emph{LLaMA} and \emph{Gemini}, few-shot experiments offer minimal improvements in semantic success.
Additionally, some practical constraints affect few-shot experimentation.
For instance, \emph{Copilot} lacks a usable API interface for automated prompting, requiring manual iterations to implement few-shot settings. 
This limitation hinders the reproducibility and scalability of few-shot evaluation across different platforms.
\emph{Baidu Qianfan} has a strict input token limit, which prevents us from providing error messages as feedback in the few-shot setting.

Overall, these findings indicate that few-shot learning is effective when the LLM possesses a baseline level of task understanding and semantic alignment, but less impactful for LLMs that lack foundational capabilities.

\section{Discussion}
\label{sec:discussion}
To explore the factors influencing LLM-supported NCDPs, we conducted experiments using a representative platform.
Our experiments examined the impact of different model choices, the language used for prompt inputs, and the role of the LLMs’ origin communities.
In this section, we discuss the implications of the findings and provide insights for future research.

\subsection{Model Selection Strategy}
With the release of a range of diverse LLMs in recent years, new opportunities for NCDP rose.
However, it is challenging to maintain an overview over model performances and select the most suitable for the use case at hand.
With a vast number of available LLMs, each exhibiting different strengths and weaknesses, choosing the right model requires careful consideration.
In this case, we select five LLMs that differ in design intent, optimization strategy, scale, and open-source or proprietary to assess their suitability for NCDPs.
According to the experiment results, the general-purpose LLMs with strong capability, such as \emph{OpenAI} models, i.e., \emph{GPT-4o} and \emph{GPT-4o-mini}, consistently show strong performance among tasks in different levels of complexity and outperform the other models.
This is true even for the LLM designed specific for software engineering tasks, e.g., \emph{Copilot}, open-source LLM optimized for efficiency and strong adaptability, e.g., \emph{LLaMA}, and LLM with advanced reasoning capability, e.g., \emph{Gemini}.
We conclude the following insights on selecting model for NCDPs.

\subsubsection{General purpose LLM suits better than coding-specific LLMs}

\emph{Copilot}, as a coding-specific model, should excel in generating functional code snippets, however, our experimental results indicate that \emph{Copilot} does not perform well in NCDPs.
Coding-specific LLMs, e.g., \emph{Copilot}, should excel in generating functional code snippets, however, our experimental results indicate that \emph{Copilot} does not perform well in NCDPs with zero-shot setting.
The primary reason for its poor performance is that generating a function for non-technical users requires more than just coding expertise, it also demands strong natural language understanding in order to interpret user prompts and project context accurately.
While coding-specific models excel at code comprehension and generation, they tend to favor technical inputs over natural language descriptions from non-technical users.
Our experiments show that \emph{Copilot}, a widely used coding-specific LLM, often produces pseudocode, incomplete code skeletons, or fails to generate functional outputs altogether indicating that it struggles to transfer natural language requirements into concrete functional implementations.
Additionally, \emph{Copilot} frequently requires further iterations with more detailed information, highlighting its limitations in natural language understanding.
Few-shot experiment results further support this observation, i.e., \emph{Copilot} starts to generate actual function logic during the few-shot round, building upon the skeletal structure it produced in the zero-shot setting.
In contrast, general-purpose LLMs balance both natural language understanding and code generation, making them more suitable for NCDPs.
Our findings indicate that natural language comprehension is more critical than pure coding ability in the NCDP context, allowing general-purpose models to better interpret and implement user requirements.

\subsubsection{Limitations of Lightweight LLMs for Complex Tasks}
While general-purpose LLMs demonstrate superior performance in NCDPs, the choice of model size and capability impacts task effectiveness.
For simpler tasks, even a lightweight model, i.e., \emph{GPT-4o-mini}, performs well, efficiently handling straightforward logic and basic automation.
Its lower computational cost makes it a viable option for quick iterations and low-complexity scenarios.
However, as task complexity increases, although the generated function remains relatively short, the lightweight models struggle to maintain accuracy and coherence. 
Complex workflows and nuanced logic require deeper contextual understanding and more robust reasoning abilities where requires a more powerful model to consistently deliver accurate and reliable results.

\subsubsection{Performance Variations Among General-Purpose LLMs}
Our experiment evaluates different general-purpose LLMs, i.e., \emph{OpenAI} models, \emph{Gemini}, and \emph{LLaMA}, using the same structured prompt.
The results reveal significant differences in their responses and only \emph{OpenAI} models prove feasible for NCDPs.
The other models can generate lengthy but invaluable outputs, i.e., \emph{Gemini} tends to produce excessive additional code snippets beyond the requested function.
While this broader output can be useful, for our experiment, it introduces unnecessary complexity, incurs unintended errors or warnings, and leads to higher token consumption.
Rather than improving the final code generation, the additional content requires users to manually extract relevant portions, which demands technical expertise and makes it less suitable for NCDPs.
This behavior suggests that some general LLMs, e.g., \emph{Gemini} prioritizes completeness over precision.
Other LLMs, e.g., \emph{LLaMA}, on the other hand, places focus on analyzing project context in the prompt.
Instead of directly generating the requested function, these LLMs tend to rewrite or restructure existing reference code, often deviating from the explicit intent of the prompt.
This indicates that these LLMs prioritize context-aware code generation, which may be beneficial in some software engineering workflows but is less aligned with the needs of NCDPs.

\subsubsection{Community background of LLMs matters}
As our end user descriptions origin in Chinese, we explore performance of comparable Chinese LLMs from \emph{Alibaba}, \emph{Baidu}, and \emph{DeepSeek}.
The \emph{Alibaba Qwen} and \emph{Baidu Qianfan} models demonstrated similar performance to the \emph{OpenAI} models, while \emph{DeepSeek-r1} performed poorly in comparison.
In detail, \emph{Qianfan} slightly outperformed \emph{Qwen}, and both surpassed \emph{GPT-4o-mini}.
While \emph{GPT-4o} outperformed the Chinese LLMs in terms of semantic success, the two Chinese models demonstrated comparable performance to \emph{GPT-4o} in terms of semantic accuracy, emphasizing the influence of community alignment to the result accuracy.
The DeepSeek model generated results similar to those of \emph{Gemini} and \emph{LLaMA}, offering a comprehensive thinking process rather than directly producing the requested function.
It can be useful for developers to understand code or for educational purposes, but it is not ideal for NCDPs, where more targeted and functional code generation is required.
The results imply that although the linguistic background of LLMs influences their performance in NCDPs, the design focus of the model and precise output are the key determinant of success.

The performance of Chinese LLMs reveals that linguistic alignment can enhance LLM performance in NCDP, but cannot replace the model design focus and capacity. 
Since the user descriptions in our dataset are originally written in Chinese, we evaluated the behavior of several representative Chinese LLMs, each with distinct design priorities, ranging from large-scale language understanding to reasoning support and enterprise-oriented deployment.
Most Chinese LLMs, e.g., \emph{Alibaba Qwen} and \emph{Baidu Qianfan}, demonstrate semantic accuracy comparable to top-performing English models, suggesting that aligning linguistic context with the input prompt language offers an initial advantage in understanding user intent. 
However, this advantage does not always translate into higher semantic success. 
For example, \emph{DeepSeek-R1}, while designed to prioritize interpretability and step-by-step reasoning, underperforms significantly due to its limited ability to generate concrete and executable code. 
Instead, it tends to output general explanations or planning-oriented text, which may be valuable in educational or debugging scenarios, but fails to satisfy the direct code generation demands of NCDPs.
These observations underscore that linguistic familiarity enhances model comprehension, but the primary determinant of NCDP performance lies in the design orientation and the capacity to produce precise and functional outputs. 

\subsubsection{Response time varies among LLMs}
The runtime of query completion is another important factor in the choice of model.
While we did not conduct systematic measurements of computation time, observations suggest potential usability impacts.
\emph{Gemini} is the fastest overall with an average response time of 9.96 seconds, but exhibits a weak functional and semantic performance. 
This suggests that lower latency may be associated with limited reasoning or decoding depth in underperforming models. 
In contrast, \emph{DeepSeek-R1} shows the longest average response time with a mean of 72.34 seconds and max of 238.39 seconds, but without proportional gains in output quality.
\emph{GPT-4o} and \emph{GPT-4o-mini} achieved a more practical balance, with moderate average latency of 18.94 seconds and 16.82 seconds, respectively, together with high success rates.
\emph{Ali-Qwen-Max} has a higher latency of 28.34 seconds, which may raise deployment concerns despite the few-shot improvements.
As \emph{Copilot} lacks API support, making it incomparable in timing metrics.
While we did not systematically evaluate response time, these incidental observations underscore its potential impact on model usability, particularly for real-time or large-scale applications.
This would be an interesting aspect for future research.

\subsection{Prompt Construction Concerns}
The language used in the prompt input plays a crucial role in determining the LLM's performance in NCDPs.
We evaluate the performance of LLM-powered NCDP with original Chinese user requirements and translated English version.
We find that achieving optimal performance requires adapting prompt design when applying different LLMs to NCDPs, as each model benefits from a distinct approach to maximize performance and ensure accurate code generation.

For a general-purpose model with strong natural language understanding, e.g., \emph{GPT-4o}, the original user prompt tends to work best, as the model can capture nuanced details from the description.
For a lightweight model, e.g., \emph{GPT-4o-mini}, the optimal strategy is the opposite.
For a model with advanced reasoning capabilities that generates excessive results, e.g., \emph{Gemini}, the original user prompt is preferred as it retains more detailed information, enabling the generation of comprehensive results.
Retaining the original prompt is more effective for easy tasks, while translating the prompt into English benefits complex tasks by allowing the model to focus more on code generation rather than natural language interpretation.
For a coding-specific model, e.g., \emph{Copilot}, English prompts generally yield better results than original Chinese ones.
When provided with Chinese input, we notice that \emph{Copilot} first translates the text into English before generating code, which can dispread the focus and introduce inaccuracies.
However, with English prompts, \emph{Copilot} is more likely to generate pseudocode or incomplete functions, limiting its suitability for NCDPs.
Although different LLMs respond differently to prompt language, the overall impact on performance remains limited.
Notably, \emph{GPT-4o} consistently outperforms other models, making it the most reliable choice for NCDPs.
These findings indicate that although prompt language can influence model behavior, strong natural language understanding is the key determinant of success. 

\subsection{Differential Effectiveness of Few-Shot Learning across LLMs}
Our experiments reveal that the few-shot prompting is effective for all applied LLMs, but the reason and adoption decision varies across LLMs.
The outperforming LLMs, such as \emph{GPT-4o}, shows benefit in both syntactic and semantic factors by iterating the syntactic errors with LLMs.
While there are limited cases showing syntactic errors, their strong and balanced performance can resolve the remaining runtime errors and provide executable and semantically correct function code.
Medium-performing LLMs, e.g., \emph{GPT-4o-mini} and \emph{Ali-Qwen-Max}, exhibit more pronounced improvements.
These LLMs have sufficient understanding and coding capacity to recognize and fix earlier mistakes when provided with feedback, and the greater room for improvement leads to more substantial gains.
The effectiveness here implies that such models are under leveraged in zero-shot setting, and still retain significant adaptability through prompting.
While these models showcase substantial improvements with few-shot prompting, their semantic success results in few-shot setting mostly approaches but still falls short of the zero-shot performance of the top-performing LLMs.
This suggests that their adaptability can compensate for initial weaknesses but may not fully bridge the performance gap with stronger models.
The case of \emph{Copilot} highlights another dimension of this dynamic. 
Unlike general-purpose LLMs, \emph{Copilot} is specialized for code generation and likely trained on narrower but denser programming corpora. 
Its tendency to generate abstract pseudocode in zero-shot settings, and its reliance on few-shot settings to complete functional logic, suggests that it prioritizes template over semantic generalization.
Feedback-based few-shot experiments effectively guide this refinement process, helping such models bridge the gap between high-level structure and executable logic.
In contrast, LLMs with overall weak performance in the zero-shot setting, such as \emph{LLaMA} and \emph{Gemini}, demonstrate minimal improvement in few-shot experiments.
While some syntactic errors can be addressed, their limited ability to generalize user intent prevents them from making meaningful semantic progress, indicating their unsuitability for NCDPs.

These results highlight that the practical adoption of few-shot prompting in NCDPs requires careful consideration. 
Although few-shot prompting can enhance syntactic and semantic success when runtime error based iteration is feasible, it comes with additional costs in terms of token consumption, inference time, and potential financial expense. 
For models without API support, e.g., \emph{Copilot}, few-shot setups may also require manual and additional engineering effort, making them less practical in large-scale deployment.
Therefore, the adoption of few-shot prompting should consider three factors: (i) the LLMs inherent capabilities in both syntax and semantics, (ii) the complexity and tolerance for errors of the target NCDP tasks, and (iii) the available computational or operational resources. 
In settings where high semantic accuracy is critical and feedback-based iteration is feasible, few-shot prompting offers clear advantages.
However, in resource-constrained environments or for models with limited generalization ability, the trade-offs may outweigh the benefits.

\subsection{Threats to Validity}
To contextualize the implications of our findings on the suitability of LLMs in NCDPs, we reflect on several factors that may affect the validity of our study. 
In particular, we discuss potential limitations related to the applied dataset (\cref{sec:threats:dataset}), the experimental platform (\cref{sec:threats:platform}), the selected use case (\cref{sec:threats:use-case}), and the long-term relevance of our observations amid ongoing LLM advancements (\cref{sec:threats:long-term-impacts}).
\label{sec:threats}
\subsubsection{Applied Dataset}
\label{sec:threats:dataset}
We use an existing dataset containing answers from 26 real users, each providing descriptions for four smart home automation tasks of varying complexity, resulting in 104 task instances in total. 
While a larger dataset could improve statistical generalizability, our study is exploratory and qualitative in nature, aiming to derive practically generalizable insights into the factors affecting LLM suitability in NCDP.
Moreover, the applied dataset is carefully designed to reflect how real non-technical users instruct LLM according to tasks of varying complexity, which aligns well with our research focus.
Therefore, we believe that this dataset is both sufficiently large to support our conclusions and well-suited to the objectives of our study.

\subsubsection{Base Platform Choice}
\label{sec:threats:platform}
To ensure a controlled environment and isolate LLM-specific effects, we adopt \emph{LLM4FaaS} as the base platform for evaluation. 
Notably, our study focuses on understanding the factors that influence LLM suitability in LLM-powered NCDPs, which the derived insights can be more broadly applicable than comparing specific platforms.
We select LLM4FaaS because of its clean architecture, i.e., functional logic generation is the responsibility of the LLM, while infrastructure abstraction is achieved through FaaS.
This design avoids additional operations or steps to achieve a reliable performance, ensuring that any performance changes resulting from modifications to LLM-related parameters can be attributed to the LLM itself.
Therefore, as an end-user-oriented, LLM-powered NCDP, LLM4FaaS aligns with the goals of our study by providing a controlled and representative environment to isolate LLM-specific behaviors and derive generalizable insights.
For those interested in platform-level comparisons and design details, we direct them to the LLM4FaaS paper~\cite{llm4faas}.

\subsubsection{Use Case Choice}
\label{sec:threats:use-case}
Our findings are based on the smart home use case, which we believe that this domain fits well for deriving broadly applicable insights into LLM performance.
It presents realistic, diverse, and often ambiguous user requirements which is an ideal testbed for evaluating how LLMs interpret and execute natural language inputs in real-world scenarios.
Comparing to domains that offer focused scenarios, e.g., chatbots or simple workflows, smart home automation encompasses a wider variety of intents, including scheduling, exception handling, and device integration, reflecting the kinds of requirements often seen in end-user development contexts. 
While domain-specific context exists, the core challenge of translating user intent into functional logic is broadly applicable.
Thus, smart home automation provides a representative and meaningful use case for evaluating the factors that influence LLM suitability in NCDPs.

\subsubsection{Long-Term Relevance of LLM Impacts}
\label{sec:threats:long-term-impacts}
While the capabilities of LLMs are expected to improve continuously, potentially mitigating some of the limitations identified in our evaluation, we argue that the underlying challenges related to design intent, community norms, and language-specific characteristics will likely persist.
Also, as LLM performance improves, user expectations are expected to rise, which may shift—rather than eliminate—the boundaries of current limitations.
Therefore, our insights regarding model selection remain relevant for guiding future NCDPs, particularly in contexts where human intent and domain-specific factors play a central role.

\subsection{Implications for Other Research Fields}
This work focuses on understanding the performance of LLMs in generating software for non-technical users.
Beyond the question of how well LLMs can perform these tasks and what causes their behavior, introducing LLM-based automation for user-oriented platforms raises interesting research questions outside the domain of computer systems research.

For example, our proposal will require additional research in human-computer interaction.
The usability and accessibility of LLMs in the domain of end-user oriented automation should be investigated further, focusing not just on the performance of LLMs but also investigating the perception that users have of the system.
It is an open question whether an unsuccessful request leads to frustration with the system or more interest and interaction as users are given an opportunity to refine their prompts.
Similarly, the cognitive load on users as they formulate their intentions as well as their trust in the system could be quantified.

As we advocate for empowering more users to build their own smart home automations (through LLMs rather than writing code), we believe that there could also be effects on behavior on a broader scale.
For individual users, being able to leverage the IoT even without a technical background could lead to both an increased reliance on LLM-based coding (without a desire to learn more about the underlying mechanisms) or more interest in technology as the user has positive interactions with the system.
Here, it could also be interesting to evaluate what mental models users have of the system and how it influences their behavior and digital literacy, especially when taking into account social and cultural factors.

Finally, we believe that this research area also warrants ethical considerations.
Leveraging LLMs to automate the smart home can have privacy implications as data on user behavior and homes is shared with the system and, potentially, an LLM service.
Then, there are concerns around responsibility when executing LLM-generated code, especially in a sensitive setting such as the private home.
If an IoT automation task could harm a user or third-party, e.g., by ignoring air quality sensor data or setting the heating to a dangerous level, and the user who prompted the automation does not have sufficient technical expertise to validate the LLM output, there is the question of accountability.
Lastly, there is the need for further research into bias and fairness.
A LLM may exhibit particularly bad behavior when confronted with automation tasks that do not fit its model of the world, e.g., when building automation around certain cultural practices.

\section{Conclusion}
\label{sec:conclusion}

LLM-supported NCDPs leverage the natural language understanding capabilities of LLMs to generate functional code based on user inputs, enabling seamless customization without requiring any technical expertise.
Understanding the factors influencing the performance of LLM-powered NCDPs can give insights on how to optimize the real-world development process, ensuring reliability, optimizing efficiency and enhancing the user experience.

In this paper, we aim to find the influencing factors that affect the performance of LLM-powered NCDPs, considering of LLM choice, prompt language variance, LLM linguistic training background, and an error-informed few-shot setting.
We provide valuable insights into the design and optimization of future platforms.
Specifically, model selection has the most significant impact on performance. 
A general-purpose LLM with advanced natural language understanding, which prioritizes providing concrete results over comprehensive ones, is generally preferred.
A LLM which the linguistic background aligns with input language also showcase an outperforming performance.
Additionally, the influence of prompt language varies across different LLMs and task complexity levels, which should be carefully considered in the design of NCDPs.
Specifically, for LLM with advanced natural language understanding and multilingual capabilities, e.g., \emph{GPT-4o}, the original user prompts should be kept to ensure the best performance.
While for LLMs with either in lightweight or with limit in natural language understanding capability, i.e., \emph{GPT-4o-mini} and \emph{Copilot}, adding a translation step can improve the performance.
Furthermore, incorporating an error-informed few-shot approach can improve LLM performance in NCDPs by providing task-specific feedback, particularly for coding-oriented and medium-performing models.
However, its effect remains secondary to model choice, and practical use requires weighing the additional engineering effort and resource costs against expected benefits.

\begin{acks}
    Partially funded by the \grantsponsor{BMFTR}{Bundesministerium für Forschung, Technologie und Raumfahrt (BMFTR, German Federal Ministry of Research, Technology and Space)}{https://www.bmftr.bund.de/EN/Home/home_node.html} in the scope of the Software Campus 3.0 (Technische Universit\"at Berlin) program -- \grantnum{BMFTR}{01IS23068}.
\end{acks}

\bibliographystyle{ACM-Reference-Format}
\bibliography{bibliography}


\begin{thebibliography}{38}


\ifx \showCODEN    \undefined \def \showCODEN     #1{\unskip}     \fi
\ifx \showISBNx    \undefined \def \showISBNx     #1{\unskip}     \fi
\ifx \showISBNxiii \undefined \def \showISBNxiii  #1{\unskip}     \fi
\ifx \showISSN     \undefined \def \showISSN      #1{\unskip}     \fi
\ifx \showLCCN     \undefined \def \showLCCN      #1{\unskip}     \fi
\ifx \shownote     \undefined \def \shownote      #1{#1}          \fi
\ifx \showarticletitle \undefined \def \showarticletitle #1{#1}   \fi
\ifx \showURL      \undefined \def \showURL       {\relax}        \fi
\providecommand\bibfield[2]{#2}
\providecommand\bibinfo[2]{#2}
\providecommand\natexlab[1]{#1}
\providecommand\showeprint[2][]{arXiv:#2}

\bibitem[Achiam et~al\mbox{.}(2023)]%
        {achiam2023gpt}
\bibfield{author}{\bibinfo{person}{Josh Achiam}, \bibinfo{person}{Steven Adler}, \bibinfo{person}{Sandhini Agarwal}, \bibinfo{person}{Lama Ahmad}, \bibinfo{person}{Ilge Akkaya}, \bibinfo{person}{Florencia~Leoni Aleman}, \bibinfo{person}{Diogo Almeida}, \bibinfo{person}{Janko Altenschmidt}, \bibinfo{person}{Sam Altman}, \bibinfo{person}{Shyamal Anadkat}, {et~al\mbox{.}}} \bibinfo{year}{2023}\natexlab{}.
\newblock \showarticletitle{Gpt-4 technical report}.
\newblock \bibinfo{journal}{\emph{arXiv preprint arXiv:2303.08774}} (\bibinfo{year}{2023}).
\newblock


\bibitem[{Artificial Analysis}({[n.\,d.]})]%
        {artificialanalysis}
\bibfield{author}{\bibinfo{person}{{Artificial Analysis}}.} \bibinfo{year}{[n.\,d.]}\natexlab{}.
\newblock \bibinfo{title}{LLM Leaderboards - Artificial Analysis}.
\newblock \bibinfo{howpublished}{\url{https://artificialanalysis.ai/leaderboards/models}}.
\newblock


\bibitem[Bao et~al\mbox{.}(2016)]%
        {bao2016microservice}
\bibfield{author}{\bibinfo{person}{Kaibin Bao}, \bibinfo{person}{Ingo Mauser}, \bibinfo{person}{Sebastian Kochanneck}, \bibinfo{person}{Huiwen Xu}, {and} \bibinfo{person}{Hartmut Schmeck}.} \bibinfo{year}{2016}\natexlab{}.
\newblock \showarticletitle{A microservice architecture for the intranet of things and energy in smart buildings}. In \bibinfo{booktitle}{\emph{Proceedings of the 1st International Workshop on Mashups of Things and APIs}}. \bibinfo{pages}{1--6}.
\newblock


\bibitem[Batista et~al\mbox{.}(2018)]%
        {10.1145/3286719.3286728}
\bibfield{author}{\bibinfo{person}{C\'{e}sar Batista}, \bibinfo{person}{Pedro~Victor Silva}, \bibinfo{person}{Everton Cavalcante}, \bibinfo{person}{Thais Batista}, \bibinfo{person}{Tiago Barros}, \bibinfo{person}{Claudio Takahashi}, \bibinfo{person}{Thiago Cardoso}, \bibinfo{person}{Jo\~{a}o~Alexandre Neto}, {and} \bibinfo{person}{Ramon Ribeiro}.} \bibinfo{year}{2018}\natexlab{}.
\newblock \showarticletitle{A Middleware Environment for Developing Internet of Things Applications}. In \bibinfo{booktitle}{\emph{Proceedings of the 5th Workshop on Middleware and Applications for the Internet of Things}} (Rennes, France) \emph{(\bibinfo{series}{M4IoT'18})}. \bibinfo{publisher}{Association for Computing Machinery}, \bibinfo{address}{New York, NY, USA}, \bibinfo{pages}{41–46}.
\newblock
\showISBNx{9781450361187}
\href{https://doi.org/10.1145/3286719.3286728}{doi:\nolinkurl{10.1145/3286719.3286728}}


\bibitem[Bermbach et~al\mbox{.}(2021)]%
        {paper_bermbach2021_cloud_engineering}
\bibfield{author}{\bibinfo{person}{David Bermbach}, \bibinfo{person}{Abhishek Chandra}, \bibinfo{person}{Chandra Krintz}, \bibinfo{person}{Aniruddha Gokhale}, \bibinfo{person}{Aleksander Slominski}, \bibinfo{person}{Lauritz Thamsen}, \bibinfo{person}{Everton Cavalcante}, \bibinfo{person}{Tian Guo}, \bibinfo{person}{Ivona Brandic}, {and} \bibinfo{person}{Rich Wolski}.} \bibinfo{year}{2021}\natexlab{}.
\newblock \showarticletitle{On the Future of Cloud Engineering}. In \bibinfo{booktitle}{\emph{Proceedings of the 9th {IEEE} International Conference on Cloud Engineering}} (San Francisco, CA, USA) \emph{(\bibinfo{series}{IC2E 2021})}. \bibinfo{publisher}{ACM}, \bibinfo{address}{New York, NY, USA}, \bibinfo{pages}{264--275}.
\newblock
\href{https://doi.org/10.1109/IC2E52221.2021.00044}{doi:\nolinkurl{10.1109/IC2E52221.2021.00044}}


\bibitem[Blackstock and Lea(2016)]%
        {blackstock2016fred}
\bibfield{author}{\bibinfo{person}{Michael Blackstock} {and} \bibinfo{person}{Rodger Lea}.} \bibinfo{year}{2016}\natexlab{}.
\newblock \showarticletitle{Fred: A hosted data flow platform for the iot}. In \bibinfo{booktitle}{\emph{Proceedings of the 1st International Workshop on Mashups of Things and APIs}}. \bibinfo{pages}{1--5}.
\newblock


\bibitem[Boh\'{e} et~al\mbox{.}(2021)]%
        {10.1145/3493369.3493600}
\bibfield{author}{\bibinfo{person}{Ilse Boh\'{e}}, \bibinfo{person}{Michiel Willocx}, \bibinfo{person}{Jorn Lapon}, {and} \bibinfo{person}{Vincent Naessens}.} \bibinfo{year}{2021}\natexlab{}.
\newblock \showarticletitle{Towards low-effort development of advanced IoT applications}. In \bibinfo{booktitle}{\emph{Proceedings of the 8th International Workshop on Middleware and Applications for the Internet of Things}} (Virtual Event, Canada) \emph{(\bibinfo{series}{M4IoT '21})}. \bibinfo{publisher}{Association for Computing Machinery}, \bibinfo{address}{New York, NY, USA}, \bibinfo{pages}{1–7}.
\newblock
\showISBNx{9781450391672}
\href{https://doi.org/10.1145/3493369.3493600}{doi:\nolinkurl{10.1145/3493369.3493600}}


\bibitem[Cai et~al\mbox{.}(2023)]%
        {cai2023low}
\bibfield{author}{\bibinfo{person}{Yuzhe Cai}, \bibinfo{person}{Shaoguang Mao}, \bibinfo{person}{Wenshan Wu}, \bibinfo{person}{Zehua Wang}, \bibinfo{person}{Yaobo Liang}, \bibinfo{person}{Tao Ge}, \bibinfo{person}{Chenfei Wu}, \bibinfo{person}{Wang You}, \bibinfo{person}{Ting Song}, \bibinfo{person}{Yan Xia}, {et~al\mbox{.}}} \bibinfo{year}{2023}\natexlab{}.
\newblock \showarticletitle{Low-code LLM: Graphical user interface over large language models}.
\newblock \bibinfo{journal}{\emph{arXiv preprint arXiv:2304.08103}} (\bibinfo{year}{2023}).
\newblock


\bibitem[Chen et~al\mbox{.}(2024)]%
        {chen2024llm2automl}
\bibfield{author}{\bibinfo{person}{Sihan Chen}, \bibinfo{person}{Weihong Zhai}, \bibinfo{person}{Chen Chai}, {and} \bibinfo{person}{Xiupeng Shi}.} \bibinfo{year}{2024}\natexlab{}.
\newblock \showarticletitle{LLM2AutoML: Zero-Code AutoML Framework Leveraging Large Language Models}. In \bibinfo{booktitle}{\emph{2024 International Conference on Intelligent Robotics and Automatic Control (IRAC)}}. IEEE, \bibinfo{pages}{285--290}.
\newblock


\bibitem[Chen et~al\mbox{.}(2022)]%
        {chen2022devicetalk}
\bibfield{author}{\bibinfo{person}{Whai-En Chen}, \bibinfo{person}{Yi-Bing Lin}, \bibinfo{person}{Tai-Hsiang Yen}, \bibinfo{person}{Syuan-Ru Peng}, {and} \bibinfo{person}{Yun-Wei Lin}.} \bibinfo{year}{2022}\natexlab{}.
\newblock \showarticletitle{DeviceTalk: A no-code low-code IoT device code generation}.
\newblock \bibinfo{journal}{\emph{Sensors}} \bibinfo{volume}{22}, \bibinfo{number}{13} (\bibinfo{year}{2022}), \bibinfo{pages}{4942}.
\newblock


\bibitem[Dantas et~al\mbox{.}(2019)]%
        {10.1145/3366610.3368100}
\bibfield{author}{\bibinfo{person}{Lucas Dantas}, \bibinfo{person}{Everton Cavalcante}, {and} \bibinfo{person}{Thais Batista}.} \bibinfo{year}{2019}\natexlab{}.
\newblock \showarticletitle{A Development Environment for FIWARE-based Internet of Things Applications}. In \bibinfo{booktitle}{\emph{Proceedings of the 6th International Workshop on Middleware and Applications for the Internet of Things}} (Davis, CA, USA) \emph{(\bibinfo{series}{M4IoT '19})}. \bibinfo{publisher}{Association for Computing Machinery}, \bibinfo{address}{New York, NY, USA}, \bibinfo{pages}{21–26}.
\newblock
\showISBNx{9781450370288}
\href{https://doi.org/10.1145/3366610.3368100}{doi:\nolinkurl{10.1145/3366610.3368100}}


\bibitem[Dubey et~al\mbox{.}(2024)]%
        {dubey2024llama}
\bibfield{author}{\bibinfo{person}{Abhimanyu Dubey}, \bibinfo{person}{Abhinav Jauhri}, \bibinfo{person}{Abhinav Pandey}, \bibinfo{person}{Abhishek Kadian}, \bibinfo{person}{Ahmad Al-Dahle}, \bibinfo{person}{Aiesha Letman}, \bibinfo{person}{Akhil Mathur}, \bibinfo{person}{Alan Schelten}, \bibinfo{person}{Amy Yang}, \bibinfo{person}{Angela Fan}, {et~al\mbox{.}}} \bibinfo{year}{2024}\natexlab{}.
\newblock \showarticletitle{The llama 3 herd of models}.
\newblock \bibinfo{journal}{\emph{arXiv preprint arXiv:2407.21783}} (\bibinfo{year}{2024}).
\newblock


\bibitem[El~Kamouchi et~al\mbox{.}(2023)]%
        {el2023low}
\bibfield{author}{\bibinfo{person}{Hind El~Kamouchi}, \bibinfo{person}{Mohamed Kissi}, {and} \bibinfo{person}{Omar El~Beggar}.} \bibinfo{year}{2023}\natexlab{}.
\newblock \showarticletitle{Low-code/No-code Development: A systematic literature review}. In \bibinfo{booktitle}{\emph{2023 14th International Conference on Intelligent Systems: Theories and Applications (SITA)}}. IEEE, \bibinfo{pages}{1--8}.
\newblock


\bibitem[Esashi et~al\mbox{.}(2025)]%
        {esashi2025action}
\bibfield{author}{\bibinfo{person}{Akiharu Esashi}, \bibinfo{person}{Pawissanutt Lertpongrujikorn}, \bibinfo{person}{Mohsen~Amini Salehi}, {and} \bibinfo{person}{Shinji Kato}.} \bibinfo{year}{2025}\natexlab{}.
\newblock \showarticletitle{Action Engine: Automatic Workflow Generation in FaaS}.
\newblock \bibinfo{journal}{\emph{Future Generation Computer Systems}} (\bibinfo{year}{2025}), \bibinfo{pages}{107947}.
\newblock


\bibitem[Gao et~al\mbox{.}(2024)]%
        {gao2024chatiot}
\bibfield{author}{\bibinfo{person}{Yi Gao}, \bibinfo{person}{Kaijie Xiao}, \bibinfo{person}{Fu Li}, \bibinfo{person}{Weifeng Xu}, \bibinfo{person}{Jiaming Huang}, {and} \bibinfo{person}{Wei Dong}.} \bibinfo{year}{2024}\natexlab{}.
\newblock \showarticletitle{ChatIoT: Zero-code Generation of Trigger-action Based IoT Programs}.
\newblock \bibinfo{journal}{\emph{Proceedings of the ACM on Interactive, Mobile, Wearable and Ubiquitous Technologies}} \bibinfo{volume}{8}, \bibinfo{number}{3} (\bibinfo{year}{2024}), \bibinfo{pages}{1--29}.
\newblock


\bibitem[{Google Translate}({[n.\,d.]})]%
        {googletranslate}
\bibfield{author}{\bibinfo{person}{{Google Translate}}.} \bibinfo{year}{[n.\,d.]}\natexlab{}.
\newblock \bibinfo{title}{Google Translate}.
\newblock \bibinfo{howpublished}{\url{https://translate.google.com}}.
\newblock


\bibitem[Hagel et~al\mbox{.}(2024)]%
        {hagel2024turning}
\bibfield{author}{\bibinfo{person}{Nathan Hagel}, \bibinfo{person}{Nicolas Hili}, {and} \bibinfo{person}{Didier Schwab}.} \bibinfo{year}{2024}\natexlab{}.
\newblock \showarticletitle{Turning Low-Code Development Platforms into True No-Code with LLMs}. In \bibinfo{booktitle}{\emph{Proceedings of the ACM/IEEE 27th International Conference on Model Driven Engineering Languages and Systems}}. \bibinfo{pages}{876--885}.
\newblock


\bibitem[{Hugging Face}({[n.\,d.]})]%
        {huggingfaceopenllm}
\bibfield{author}{\bibinfo{person}{{Hugging Face}}.} \bibinfo{year}{[n.\,d.]}\natexlab{}.
\newblock \bibinfo{title}{Open LLM Leaderboard}.
\newblock \bibinfo{howpublished}{\url{http://huggingface.co/open-llm-leaderboard}}.
\newblock


\bibitem[{Hugging Face and lmsys.org}({[n.\,d.]})]%
        {chatbotarena}
\bibfield{author}{\bibinfo{person}{{Hugging Face and lmsys.org}}.} \bibinfo{year}{[n.\,d.]}\natexlab{}.
\newblock \bibinfo{title}{Chatbot Arena Leaderboard}.
\newblock \bibinfo{howpublished}{\url{https://huggingface.co/spaces/lmarena-ai/chatbot-arena-leaderboard}}.
\newblock


\bibitem[K{\"o}k et~al\mbox{.}(2024)]%
        {kok2024iot}
\bibfield{author}{\bibinfo{person}{{\.I}brahim K{\"o}k}, \bibinfo{person}{Orhan Demirci}, {and} \bibinfo{person}{Suat {\"O}zdemir}.} \bibinfo{year}{2024}\natexlab{}.
\newblock \showarticletitle{When IoT Meet LLMs: Applications and Challenges}. In \bibinfo{booktitle}{\emph{2024 IEEE International Conference on Big Data (BigData)}}. IEEE, \bibinfo{pages}{7075--7084}.
\newblock


\bibitem[Koubaa et~al\mbox{.}(2025)]%
        {koubaa2025next}
\bibfield{author}{\bibinfo{person}{Anis Koubaa}, \bibinfo{person}{Adel Ammar}, {and} \bibinfo{person}{Wadii Boulila}.} \bibinfo{year}{2025}\natexlab{}.
\newblock \showarticletitle{Next-generation human-robot interaction with ChatGPT and robot operating system}.
\newblock \bibinfo{journal}{\emph{Software: Practice and Experience}} \bibinfo{volume}{55}, \bibinfo{number}{2} (\bibinfo{year}{2025}), \bibinfo{pages}{355--382}.
\newblock


\bibitem[Li et~al\mbox{.}(2025)]%
        {li2025large}
\bibfield{author}{\bibinfo{person}{Jun Li}, \bibinfo{person}{Lixian Li}, \bibinfo{person}{Jin Liu}, \bibinfo{person}{Xiao Yu}, \bibinfo{person}{Xiao Liu}, {and} \bibinfo{person}{Jacky~Wai Keung}.} \bibinfo{year}{2025}\natexlab{}.
\newblock \showarticletitle{Large language model ChatGPT versus small deep learning models for self-admitted technical debt detection: Why not together?}
\newblock \bibinfo{journal}{\emph{Software: Practice and Experience}} \bibinfo{volume}{55}, \bibinfo{number}{1} (\bibinfo{year}{2025}), \bibinfo{pages}{3--28}.
\newblock


\bibitem[Liu et~al\mbox{.}(2024)]%
        {liu2024empirical}
\bibfield{author}{\bibinfo{person}{Yongkun Liu}, \bibinfo{person}{Jiachi Chen}, \bibinfo{person}{Tingting Bi}, \bibinfo{person}{John Grundy}, \bibinfo{person}{Yanlin Wang}, \bibinfo{person}{Ting Chen}, \bibinfo{person}{Yutian Tang}, {and} \bibinfo{person}{Zibin Zheng}.} \bibinfo{year}{2024}\natexlab{}.
\newblock \showarticletitle{An Empirical Study on Low Code Programming using Traditional vs Large Language Model Support}.
\newblock \bibinfo{journal}{\emph{arXiv:2402.01156 [cs.SE]}} (\bibinfo{year}{2024}).
\newblock


\bibitem[{LLM Stats}({[n.\,d.]})]%
        {llmstats}
\bibfield{author}{\bibinfo{person}{{LLM Stats}}.} \bibinfo{year}{[n.\,d.]}\natexlab{}.
\newblock \bibinfo{title}{LLMStats.com – Model Leaderboard and Trends}.
\newblock \bibinfo{howpublished}{\url{https://llm-stats.com}}.
\newblock


\bibitem[Lu et~al\mbox{.}(2025)]%
        {lu2025performance}
\bibfield{author}{\bibinfo{person}{Zhengxian Lu}, \bibinfo{person}{Fangyu Wang}, \bibinfo{person}{Zhiwei Xu}, \bibinfo{person}{Fei Yang}, {and} \bibinfo{person}{Tao Li}.} \bibinfo{year}{2025}\natexlab{}.
\newblock \showarticletitle{On the performance and memory footprint of distributed training: An empirical study on transformers}.
\newblock \bibinfo{journal}{\emph{Software: Practice and Experience}} \bibinfo{volume}{55}, \bibinfo{number}{7} (\bibinfo{year}{2025}), \bibinfo{pages}{1266--1284}.
\newblock


\bibitem[Martins et~al\mbox{.}(2023)]%
        {martins2023combining}
\bibfield{author}{\bibinfo{person}{Jos{\'e} Martins}, \bibinfo{person}{Frederico Branco}, {and} \bibinfo{person}{Henrique Mamede}.} \bibinfo{year}{2023}\natexlab{}.
\newblock \showarticletitle{Combining low-code development with ChatGPT to novel no-code approaches: a focus-group study}.
\newblock \bibinfo{journal}{\emph{Intelligent Systems with Applications}}  \bibinfo{volume}{20} (\bibinfo{year}{2023}), \bibinfo{pages}{200289}.
\newblock


\bibitem[Monteiro et~al\mbox{.}(2025)]%
        {monteiro2025nocodegpt}
\bibfield{author}{\bibinfo{person}{Mauricio Monteiro}, \bibinfo{person}{Bruno~Castelo Branco}, \bibinfo{person}{Samuel Silvestre}, \bibinfo{person}{Guilherme Avelino}, {and} \bibinfo{person}{Marco~Tulio Valente}.} \bibinfo{year}{2025}\natexlab{}.
\newblock \showarticletitle{NoCodeGPT: A No-Code Interface for Building Web Apps With Language Models}.
\newblock \bibinfo{journal}{\emph{Software: Practice and Experience}} (\bibinfo{year}{2025}).
\newblock


\bibitem[Paliwal et~al\mbox{.}(2024)]%
        {paliwal2024low}
\bibfield{author}{\bibinfo{person}{Gunjan Paliwal}, \bibinfo{person}{Anujkumarsinh Donvir}, \bibinfo{person}{Praveen Gujar}, {and} \bibinfo{person}{Sriram Panyam}.} \bibinfo{year}{2024}\natexlab{}.
\newblock \showarticletitle{Low-Code/No-Code Meets GenAI: A New Era in Product Development}. In \bibinfo{booktitle}{\emph{2024 IEEE Eighth Ecuador Technical Chapters Meeting (ETCM)}}. IEEE, \bibinfo{pages}{1--9}.
\newblock


\bibitem[Petrukha et~al\mbox{.}(2025)]%
        {petrukha2025swifteval}
\bibfield{author}{\bibinfo{person}{Ivan Petrukha}, \bibinfo{person}{Yana Kurliak}, {and} \bibinfo{person}{Nataliia Stulova}.} \bibinfo{year}{2025}\natexlab{}.
\newblock \showarticletitle{SwiftEval: Developing a Language-Specific Benchmark for LLM-generated Code Evaluation}. In \bibinfo{booktitle}{\emph{2025 IEEE/ACM Second International Conference on AI Foundation Models and Software Engineering (Forge)}}. IEEE, \bibinfo{pages}{73--77}.
\newblock


\bibitem[Pfandzelter and Bermbach(2019)]%
        {paper_pfandzelter2019_functions_vs_streams}
\bibfield{author}{\bibinfo{person}{Tobias Pfandzelter} {and} \bibinfo{person}{David Bermbach}.} \bibinfo{year}{2019}\natexlab{}.
\newblock \showarticletitle{IoT Data Processing in the Fog: Functions, Streams, or Batch Processing?}. In \bibinfo{booktitle}{\emph{Proceedings of the 1st Workshop on Efficient Data Movement in Fog Computing}} (Prague, Czech Republic) \emph{(\bibinfo{series}{DaMove 2019})}. \bibinfo{publisher}{IEEE}, \bibinfo{address}{New York, NY, USA}, \bibinfo{pages}{201--206}.
\newblock
\href{https://doi.org/10.1109/ICFC.2019.00033}{doi:\nolinkurl{10.1109/ICFC.2019.00033}}


\bibitem[Sufi(2023)]%
        {sufi2023algorithms}
\bibfield{author}{\bibinfo{person}{Fahim Sufi}.} \bibinfo{year}{2023}\natexlab{}.
\newblock \showarticletitle{Algorithms in low-code-no-code for research applications: A practical review}.
\newblock \bibinfo{journal}{\emph{Algorithms}} \bibinfo{volume}{16}, \bibinfo{number}{2} (\bibinfo{year}{2023}), \bibinfo{pages}{108}.
\newblock


\bibitem[Team et~al\mbox{.}(2023)]%
        {team2023gemini}
\bibfield{author}{\bibinfo{person}{Gemini Team}, \bibinfo{person}{Rohan Anil}, \bibinfo{person}{Sebastian Borgeaud}, \bibinfo{person}{Jean-Baptiste Alayrac}, \bibinfo{person}{Jiahui Yu}, \bibinfo{person}{Radu Soricut}, \bibinfo{person}{Johan Schalkwyk}, \bibinfo{person}{Andrew~M Dai}, \bibinfo{person}{Anja Hauth}, \bibinfo{person}{Katie Millican}, {et~al\mbox{.}}} \bibinfo{year}{2023}\natexlab{}.
\newblock \showarticletitle{Gemini: a family of highly capable multimodal models}.
\newblock \bibinfo{journal}{\emph{arXiv preprint arXiv:2312.11805}} (\bibinfo{year}{2023}).
\newblock


\bibitem[Ulfsnes et~al\mbox{.}(2024)]%
        {ulfsnes2024transforming}
\bibfield{author}{\bibinfo{person}{Rasmus Ulfsnes}, \bibinfo{person}{Nils~Brede Moe}, \bibinfo{person}{Viktoria Stray}, {and} \bibinfo{person}{Marianne Skarpen}.} \bibinfo{year}{2024}\natexlab{}.
\newblock \showarticletitle{Transforming software development with generative AI: empirical insights on collaboration and workflow}.
\newblock In \bibinfo{booktitle}{\emph{Generative AI for effective software development}}. \bibinfo{publisher}{Springer}, \bibinfo{pages}{219--234}.
\newblock


\bibitem[Upadhyaya(2023)]%
        {upadhyaya2023low}
\bibfield{author}{\bibinfo{person}{Nitesh Upadhyaya}.} \bibinfo{year}{2023}\natexlab{}.
\newblock \showarticletitle{Low-Code/No-Code Platforms and Their Impact on Traditional Software Development: A Literature Review}.
\newblock \bibinfo{journal}{\emph{No-Code Platforms and Their Impact on Traditional Software Development: A Literature Review (March 21, 2023)}} (\bibinfo{year}{2023}).
\newblock


\bibitem[Wang et~al\mbox{.}({[n.\,d.]})]%
        {llm4faasdataset}
\bibfield{author}{\bibinfo{person}{Minghe Wang}, \bibinfo{person}{Tobias Pfandzelter}, \bibinfo{person}{Trever Schirmer}, {and} \bibinfo{person}{David Bermbach}.} \bibinfo{year}{[n.\,d.]}\natexlab{}.
\newblock \bibinfo{title}{LLM4FaaS Dataset}.
\newblock \bibinfo{howpublished}{\url{https://github.com/Mhwwww/LLM4FaaS-dataset}}.
\newblock


\bibitem[Wang et~al\mbox{.}(2025)]%
        {llm4faas}
\bibfield{author}{\bibinfo{person}{Minghe Wang}, \bibinfo{person}{Tobias Pfandzelter}, \bibinfo{person}{Trever Schirmer}, {and} \bibinfo{person}{David Bermbach}.} \bibinfo{year}{2025}\natexlab{}.
\newblock \showarticletitle{LLM4FaaS: No-Code Application Development using LLMs and FaaS}.
\newblock \bibinfo{journal}{\emph{arXiv preprint arXiv:2502.14450}} (\bibinfo{year}{2025}).
\newblock


\bibitem[Werner et~al\mbox{.}(2017)]%
        {paper_werner2017_openhab}
\bibfield{author}{\bibinfo{person}{Sebastian Werner}, \bibinfo{person}{Frank Pallas}, {and} \bibinfo{person}{David Bermbach}.} \bibinfo{year}{2017}\natexlab{}.
\newblock \showarticletitle{Designing Suitable Access Control for Web-Connected Smart Home Platforms}. In \bibinfo{booktitle}{\emph{Proceedings of the 13th International Workshop on Engineering Service-Oriented Applications and Cloud Services}} (Malaga, Spain) \emph{(\bibinfo{series}{WESOACS 2017})}. \bibinfo{publisher}{Springer}, \bibinfo{address}{Cham, Switzerland}, \bibinfo{pages}{240--251}.
\newblock
\href{https://doi.org/10.1007/978-3-319-91764-1_19}{doi:\nolinkurl{10.1007/978-3-319-91764-1_19}}


\bibitem[Younes et~al\mbox{.}(2018)]%
        {10.1145/3286719.3286725}
\bibfield{author}{\bibinfo{person}{Walid Younes}, \bibinfo{person}{Sylvie Trouilhet}, \bibinfo{person}{Fran\c{c}oise Adreit}, {and} \bibinfo{person}{Jean-Paul Arcangeli}.} \bibinfo{year}{2018}\natexlab{}.
\newblock \showarticletitle{Towards an Intelligent User-Oriented Middleware for Opportunistic Composition of Services in Ambient Spaces}. In \bibinfo{booktitle}{\emph{Proceedings of the 5th Workshop on Middleware and Applications for the Internet of Things}} (Rennes, France) \emph{(\bibinfo{series}{M4IoT'18})}. \bibinfo{publisher}{Association for Computing Machinery}, \bibinfo{address}{New York, NY, USA}, \bibinfo{pages}{25–30}.
\newblock
\showISBNx{9781450361187}
\href{https://doi.org/10.1145/3286719.3286725}{doi:\nolinkurl{10.1145/3286719.3286725}}


\end{thebibliography}

\end{document}